\documentclass[%
reprint,
superscriptaddress,
 amsmath,amssymb,
 aps,
]{revtex4-2}

\usepackage{graphicx}
\usepackage{siunitx}
\usepackage{dcolumn}
\usepackage{bm}
\usepackage{hyperref}
\usepackage[mathlines]{lineno}

\begin{document}

\title{\textbf{Optimal mechanical interactions direct multicellular network formation on elastic substrates}}

\author{Patrick S. Noerr}

\affiliation{Department of Physics, University of California Merced, CA 95343, USA
}

\author{Jose E. Zamora Alvarado}

\affiliation{Department of Materials and Biomaterials Science and Engineering , University of California Merced, CA 95343, USA}

\author{Farnaz Golnaraghi}%

 \affiliation{Department of Physics, University of California Merced, CA 95343, USA
}

\author{Kara E. McCloskey}

\affiliation{Department of Materials and Biomaterials Science and Engineering , University of California Merced, CA 95343, USA}

\author{Ajay Gopinathan}
\thanks{Correspondence: agopinathan@ucmerced.edu}
 \affiliation{Department of Physics, University of California Merced, CA 95343, USA
}%

\author{Kinjal Dasbiswas}
\thanks{Correspondence: kdasbiswas@ucmerced.edu}
\affiliation{%
 Department of Physics, University of California Merced, CA 95343, USA
}%


\begin{abstract}
Cells self-organize into functional, ordered structures during tissue morphogenesis, a process that is evocative of colloidal self-assembly into engineered soft materials. Understanding how inter-cellular mechanical interactions may drive the formation of  ordered and functional multicellular structures is important in developmental biology and tissue engineering. Here, by combining an agent-based model for contractile cells on elastic substrates with  endothelial cell culture experiments, we show that substrate deformation-mediated mechanical interactions between cells can cluster and align them into branched networks. Motivated by the structure and function of vasculogenic networks, we predict how measures of network connectivity like percolation and fractal dimension, as well as local morphological features including junctions, branches, and rings depend on  cell contractility and density, and on substrate elastic properties including stiffness and compressibility. We predict and confirm with experiments that cell network formation is substrate stiffness-dependent, being optimal at intermediate stiffness. Overall, we show that long-range, mechanical interactions provide an optimal  and general strategy for multi-cellular self-organization, leading to more robust and efficient realization of space-spanning networks than through just local inter-cellular interactions.
\begin{description}
\item[Keywords]
\small{Mechanobiology, Soft Matter, Agent-based modeling, Cell culture experiments, Multicellular networks} 
\end{description}
\end{abstract}

\maketitle


\section*{Introduction}
The morphogenesis of biological tissue involves the organization of cells into functional, self-assembled structures \cite{Forgacs2005}. The aggregation of cells into ordered structures 
requires effectively attractive cell-cell interactions \cite{foty2005differential}. An example that is significant for biological development, disease and tissue engineering, is the morphogenesis of blood vessels. This is initiated by patterned structures of endothelial cells (ECs), which align end to end to form elongated chains that intersect to give a branched morphology.  Although the conditions required for vascular-like development in engineered \textit{in vitro} systems are well established and EC vascular networks have been  mathematically modeled using various approaches \cite{Manoussaki1996,Gamba2003,Szabo2007,vanOers2014, Kleinstreuer2013, Ramos2018, Stepanova2021}, the nature of the cell-cell interactions that drive the ECs to find each other to form networks, and the dependence of these interactions on matrix stiffness, have not been definitively identified.

 The emergence of complex structures from the interactions of individual agents bears resemblance to colloidal self-assembly.  For example, dipolar particles, such as ferromagnetic colloids, will align end-to-end into equilibrium, linear structures such as chains or rings \cite{deGennes1970}. At higher densities, the chains intersect to form gel-like network structures \cite{Ilg2011}.  Such structures have been studied in simulation in the context of active dipoles representing synthetic active colloids endowed with a permanent or induced dipole moment \cite{Andreas2015, Liao2020, Sakai2020}, and swimming microorganisms \cite{Guzman-Lastra2016} such as magnetotactic bacteria \cite{Telezki2020}. Animal cells that adhere to and crawl on elastic substrates and interact through mechanical deformations of the substrate \cite{safran_13} are also expected to attract and align to form multicellular structures \cite{Bischofs2003}. Such mechanically directed self-organization of cells into functional structures such as vascular networks imply that network morphology depends on substrate stiffness.



While cells routinely communicate using chemical signals, they also sense each other through mechanical forces that they exert on each other, either through direct cell-cell contacts or indirectly, through mutual deformations of a compliant, extracellular substrate \cite{pelham_97, dembo_99}. 
Large and measurable substrate deformations \cite{balaban_01} are produced by several cell types that use mechanical forces actively generated by myosin motors in their actin cytoskeleton to change shape, move, and sense their surroundings \cite{gardel_15}. Elastic substrate-mediated intercellular mechanical communication  has been demonstrated for several contractile cell types. Endothelial cells modulate their intercellular contact frequency according to substrate stiffness \cite{ Reinhart2008}, cardiomyocytes synchronize their beating with substrate mechanical oscillations induced by a distant probe \cite{Tang11, Nitsan2016}, and fibroblasts interact at long range through their structural remodeling of fibrous extracellular media \cite{Notbohm2015, abhilash_14}.

\begin{figure*}[hbt!]  
		\centering
		\includegraphics[width=\linewidth]{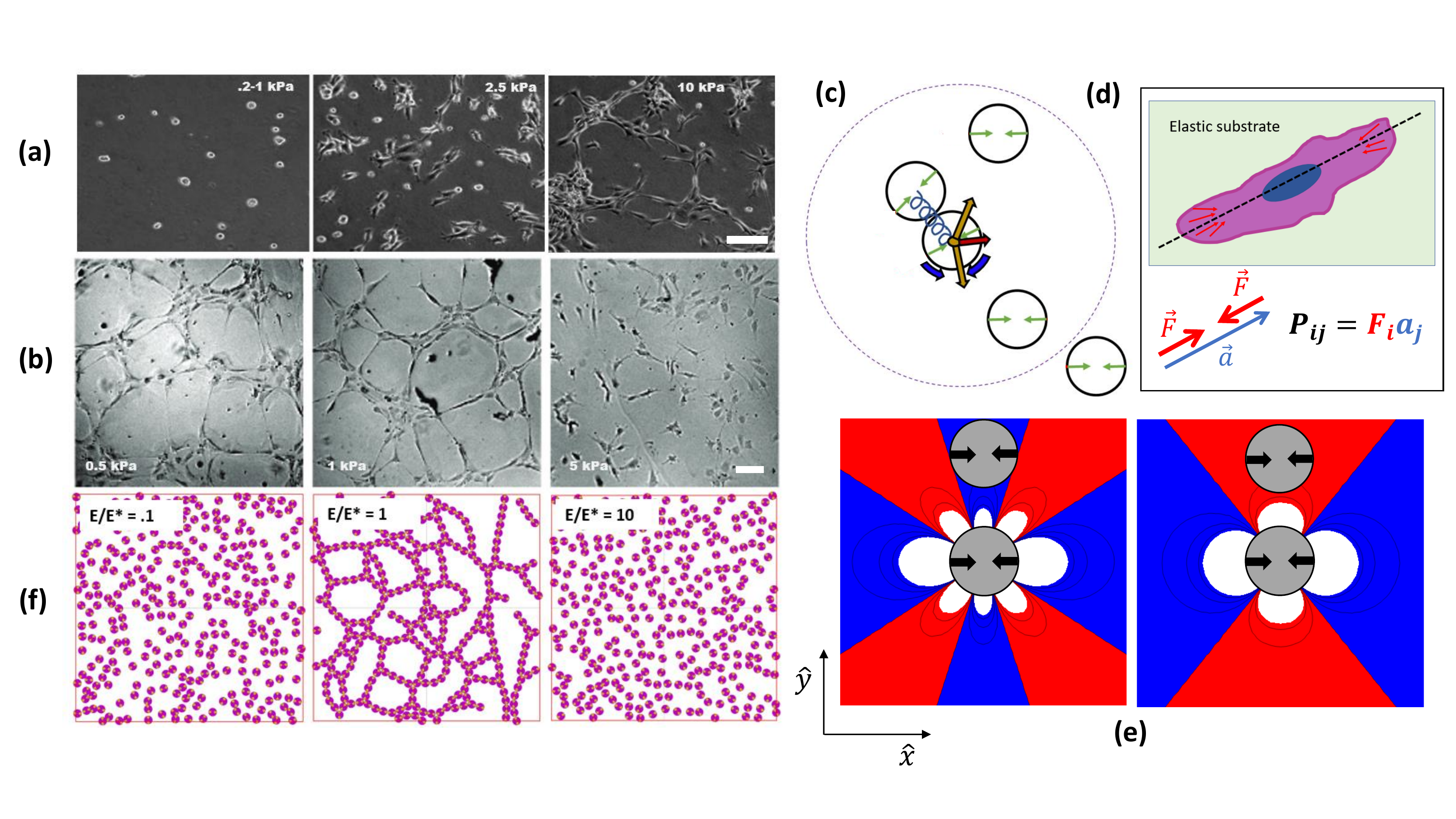}
		\caption{Cell network formation is optimized by substrate stiffness. (a) 
		Bovine aortic endothelial cells (BAECs) cultured on polyacrylamide substrates of varying stiffness that were coated with collagen. For the  collagen concentration shown here (1 $\mu$g/ml), the cells adhered to the substrate but did not form networks at lower substrate stiffness (0.2-1 kPa), but they did so at higher stiffness (2.5  kPa and higher). Images from Ref.~\cite{Califano2008}. (b)  Human umbilical vascular endothelial cells (HUVECs) cultured on polyacrylamide hydrogel substrates of varying stiffness that were coated with matrigel. At high stiffness (5 kPa and glass), the cells did not form networks, but did so on softer substrates (0.5 and 1 kPa). Scale bars = $100 \: \mu \mathrm{m}$. Images from Ref.~\cite{Rudiger2020}. (c) Cartoon of a simulation snapshot where green arrows indicate the cell's force dipole, large purple dashed ring denotes the elastic interaction range, blue squiggle indicates a repulsive spring to prevent overlap, bold gold arrows represent force vectors due to elastic interactions, bold red arrows represents the net force vector on the central cell, bold blue arrow represents torque on central cell due to elastic interaction with neighbors. (d) Cartoon cell deforming the surrounding elastic substrate by applying forces along a main axis.(e) $u_{xx}$ component of the strain field caused by a contractile force dipole centered at the origin pinching along the x-axis for $\nu$ = 0.5(left) and $\nu$ = 0.1(right) with coordinate axes labeled. (f) Simulation snapshots of 300 cells modeled as contractile force dipoles that move and reorient according to substrate-mediated cell-cell elastic interaction forces. Cells form percolating networks only for a range of substrate stiffness values centered around an optimal stiffness, $E^{*}$, above which cells exert maximal traction force. For substrates around optimal stiffness ($E/E^{\ast} \sim 1$), the substrate-mediated cell--cell elastic interactions are maximal and can be much larger than the noise in cell movements, whereas for very soft ($E/E^{\ast} \ll 1$) or very stiff ($E/E^{\ast} \gg 1$) substrates, the elastic interactions are likely to be overwhelmed by  noise, resulting in a lack of ordered structures. $ E^{\ast} = 1$ kPa was the optimal stiffness value used in these simulations.}
		\label{fig:MasterPlot2} 
	\end{figure*}

Cells sense substrate mechanical deformations through mechanotransduction occurring at the biomolecular scale \cite{discher_05}. Such cellular signaling is carried out by proteins associated with the cell--substrate adhesions, that are in turn connected to the cell's cytoskeletal force-generating machinery \cite{balaban_01}. At a coarse-grained level, the contractile apparatus of cells adhered to an extracellular substrate  can be modeled as active elastic inclusions \cite{Schwarz2002}, which adapts the theory of material inclusions developed by Eshelby \cite{Eshelby1957}, to describe cellular contractility as force dipoles embedded in an elastic medium.  This general theoretical approach  predicts how multicellular and subcellular cytoskeletal organization depend on substrate stiffness \cite{Bischofs2003, Bischofs2005}. It has been applied successfully to explain experimental observations of substrate stiffness-dependent structural order in a variety of cell types in a unified manner \cite{zemel_10, engler_08, friedrich_11, dasbiswas_15, dasbiswas_18}. While these previous works focused on the stationary configurations of elastic dipoles in the context of adherent cells \cite{Bischofs2004, Bischofs2006}, we now consider cell self-assembly when the cellular dipoles are free to translate and rotate in response to mechanical forces, thereby serving as minimal models for contractile cells that adhere to, spread and and crawl on soft media.   We show that cell-cell mechanical interactions mediated by a compliant elastic substrate can drive network formation, and that the resulting network morphology  is inherently sensitive to substrate stiffness. 

Coarse-grained material properties of the cellular micro-environment, such as its stiffness and viscosity, are known to play crucial roles in determining cell structure and function \cite{Discher2005, Engler2006, Chaudhuri2020}.  ~\hyperref[fig:MasterPlot2]{Figs. 1 a, b} are taken from two different cell culture experiments on substrates of varying stiffness.  In the first experiment~(\hyperref[fig:MasterPlot2]{Fig. 1 a}), it was shown that, under certain conditions, bovine endothelial cells formed networks preferentially on stiffer substrates ($E \sim 10$ KPa) \cite{Califano2008}, 
More recently(~\hyperref[fig:MasterPlot2]{Fig. 1 b}), it was shown that human umbilical vascular endothelial cells (HUVECs) assemble into networks on softer substrates ($E \sim 1$ kPa) but fail to do so on stiffer substrates, independently of the type of hydrogel used \cite{Rudiger2020}.  Both these experiments show that EC network formation is sensitive to substrate stiffness, and therefore suggest that cell mechanical interactions mediated by the substrate are involved.

\section*{Model and Results}
\subsection*{ Substrate stiffness-dependent endothelial cell network organization motivates model for cell mechanical interactions.}


To model cell network formation, we incorporate substrate-mediated cell mechanical interactions into an agent-based model for cell motility \cite{Copenhagen2018}. This captures the dynamic re-arrangements of cells into favorable configurations.  In our agent-based approach \cite{Bose2021, BoseNoerr2022}, summarized in \hyperref[fig:MasterPlot2]{Fig. 1 c}, we consider a system of $N$ particles, each a disk of diameter $\sigma$. Depending on the context, each disk could model a cell or its constituent parts, and their motion represent both cell migration as well as cell spreading or shape change dynamics.  Details of the cell shape are not included in this minimal model. These model cells self-organize according to substrate friction--dominated overdamped dynamics that depend on inter-cell interactions, as well as individual cell stochastic movements described by an effective diffusion. The model incorporates both short-range, steric and long-range, substrate-mediated elastic interactions between cells, and is detailed in the Methods section 


The ubiquitous traction force pattern generated by a single polarized cell with a long axis ${\mathbf a}$ and exerting a typical force $\mathbf {F}$ at its adhesions, can be modeled as a force dipole, $P_{ij} = F_{i} a_{j}$ (\hyperref[fig:MasterPlot2]{Fig. 1 d}). Note that the cell traction forces are generated by actomyosin units within the cell, each of which acts as a force dipole. Therefore, the disks in our model simulations could represent parts of a cell, and their motion represent the dynamics of cell protrusions. The resulting deformation induced by a force dipole in the elastic substrate is given by the strain, $u_{ij}$, which is determined by a force balance in linear elastic theory (see SI section A), and depends on the material properties of the elastic medium, specifically, the stiffness or Young's modulus $E$, and the compressibility, given by the Poisson's ratio $\nu$ \cite{landau_lifshitz_elasticity}. The substrate deformation ($u_{xx}$ component of strain) generated by a dipole (oriented along the laboratory $x-$axis) embedded on the surface of a linear elastic medium is shown in (\hyperref[fig:MasterPlot2]{Fig. 1 e})) for two representative values of $\nu$.  Here, the blue (red) coloring represents expanded (compressed) regions of the substrate. 

A second contractile force dipole will tend to position itself in and align its axis along the local principal stretch in the medium to reduce the substrate deformation. The resulting interaction potential arises from the minimal coupling of one dipole (denoted by $\beta$) with the medium strain induced by the other (denoted by $\alpha$), and is given by $W^{\alpha \beta} = P^{\beta}_{ij} u^{\alpha}_{ij}$\cite{safran_13}.  The interaction energy between two dipoles then decays with their separation distance as $W^{\alpha \beta} \sim (P^{2}/E) \cdot r_{\alpha \beta}^{-3}$. We denote the characteristic elastic interaction energy when the dipoles are separated by only one cell length as, $\mathcal{E}_{c} = P^{2}/(16 E \sigma^{3})$, where the detailed expression is derived in the SI Section A. This coarse-grained description abstracts out the biophysical details of mechanotransduction, but provides a simple physical model for the cell response to deformations in their elastic medium \cite{Bischofs2003}.

Representative simulation snapshots (\hyperref[fig:MasterPlot2]{Fig. 1 f}) of final configurations show that elastic dipolar interactions induce network formation in a stiffness-dependent manner. The central snapshot corresponds to an optimal substrate stiffness $E^{\ast}$ at which elastic interactions are maximal, while those to the left (right) correspond to substrates that are too soft (stiff) for connected network formation.  The origin of this optimal  stiffness lies in the adaptation of cell contractile forces to their substrate stiffness, as we discuss later.

\begin{figure*}[hbt!]  
	\centering
	\includegraphics[width=1\linewidth]{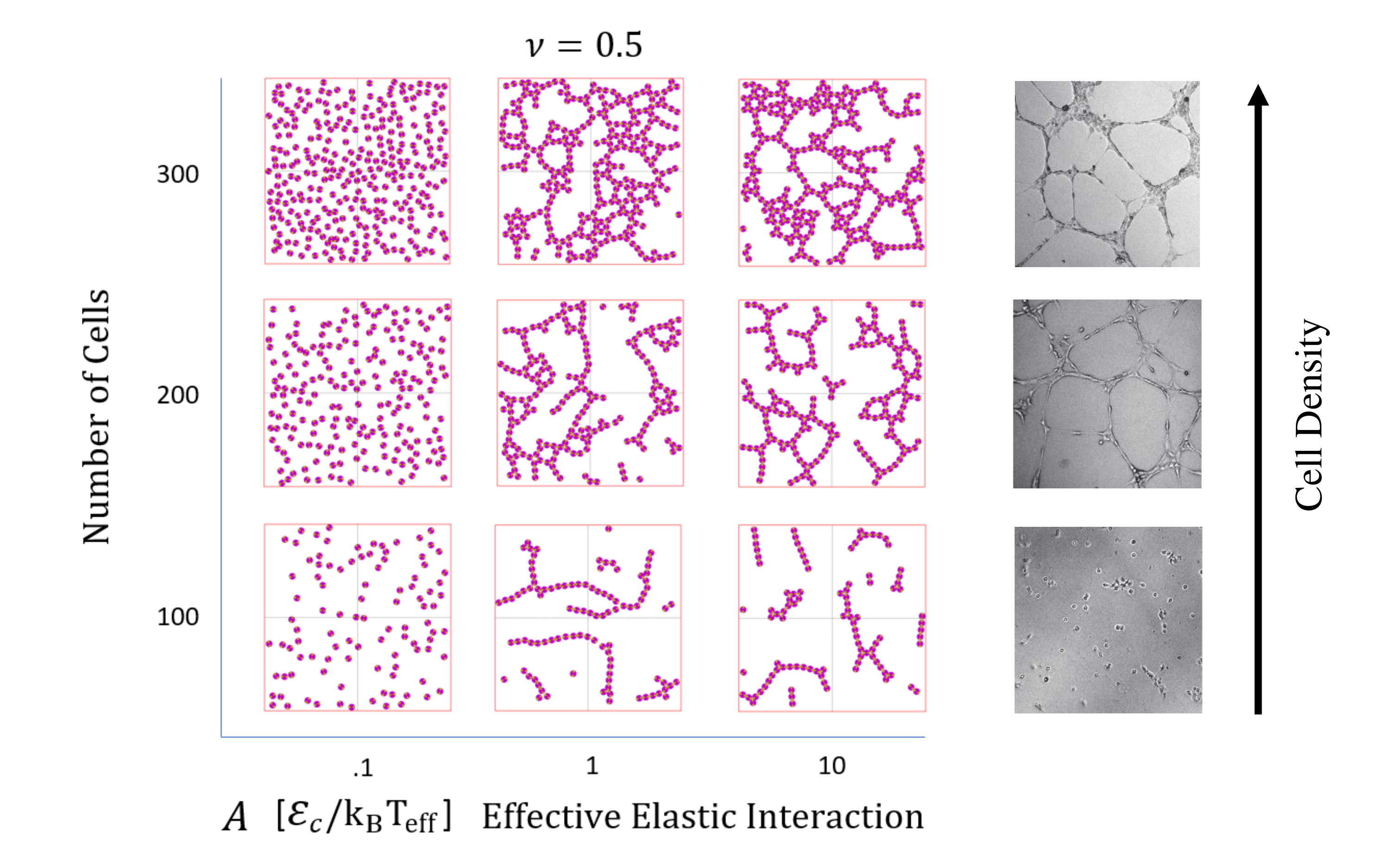}
	\caption{Simulation snapshots showing representative final configurations of model cell dipoles. We explore the parameter space of number of cells and $A \equiv \tfrac{\mathcal{E}_{c}}{k_\mathrm{B}T_{\mathrm{eff}}}\:$, the ratio of the characteristic elastic interaction strength and noise, for Poisson's ratio, $\nu = 0.5$. At lower packing fractions, cells form disconnected linear clusters. At lower $A$ values, cells  remain isolated, but at moderate values of $A$ and sufficient packing fraction, cells form space spanning  network configurations characterized by rings, branches, and junctions. At higher packing fractions, clumpy structures such as what previous literature calls "4-rings" occur frequently \cite{Bischofs2006}. The tendency for cells to form only local connections at low packing fraction and form space spanning structures at higher packing fraction is consistent with experimental images of endothelial cells cultured on hydrogel substrates (right column; images reproduced from Ref.~\cite{Rudiger2020}).}
	\label{fig:Figure2} 
	\end{figure*}

\begin{figure*}[hbt!]  
	\centering
	\includegraphics[width=1\linewidth]{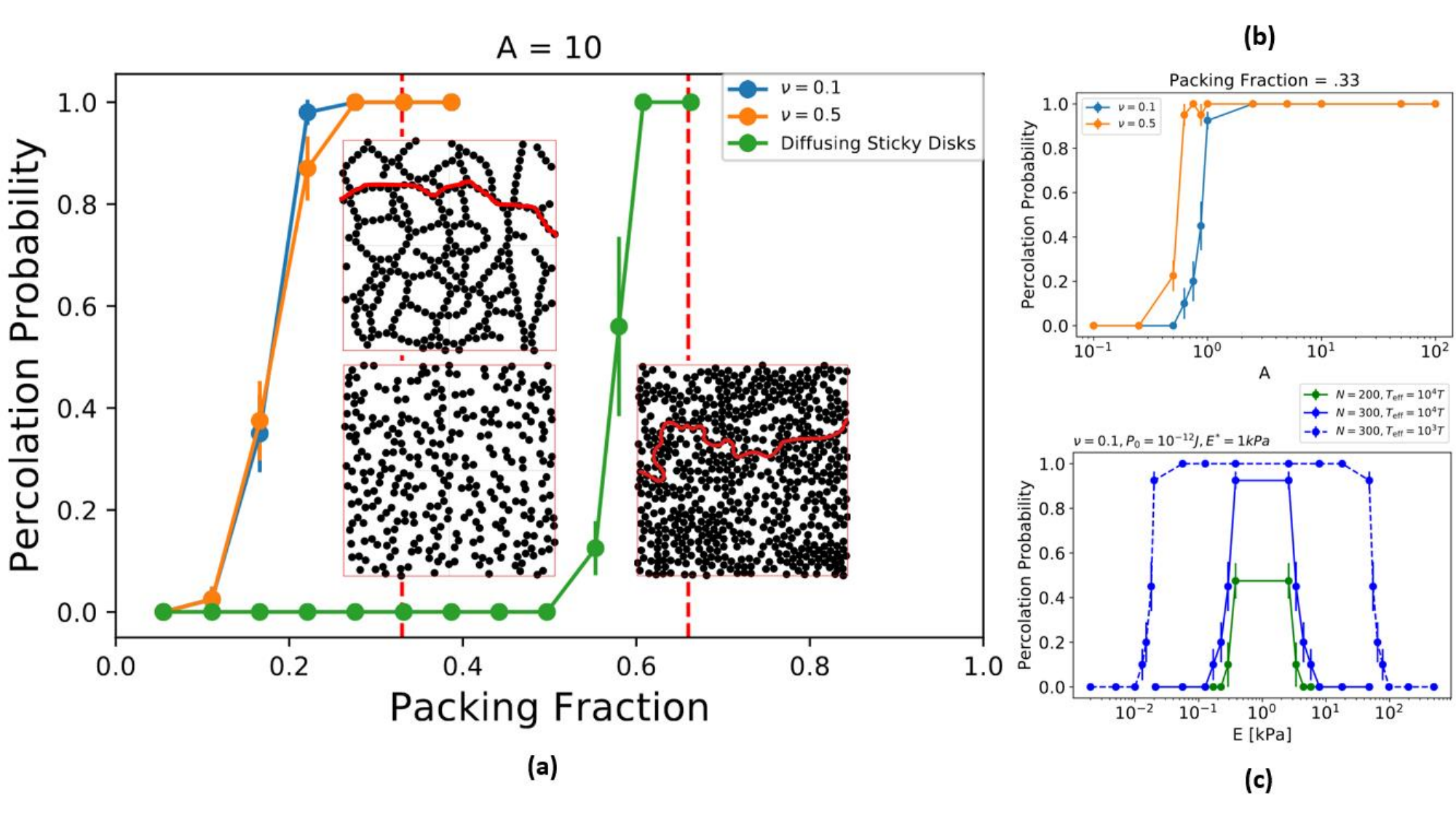}
	\caption{
 Analysis of connectivity percolation of model cell networks formed by substrate-mediated elastic interactions predicts network dependence on substrate stiffness.
 (a) Percolation probability for elastic dipoles - blue and orange - and diffusing sticky disks - green - as a function of area fraction. Elastic dipoles undergo the percolation transition at lower packing fractions than purely diffusive, sticky disks. The insets show characteristic final configurations for both elastic dipoles and sticky disks at a packing fraction of $.33(N=300)$, with an example percolating path shown in red. (b) The percolation transition is dependent on the effective elastic interaction. (c) Percolation probability as a function of elastic substrate stiffness where the optimal stiffness is assumed to be 1 kPa. Percolation peak is centered on critical stiffness and has a width dependent on both packing fraction and effective temperature.}
	\label{fig:Figure3} 
	\end{figure*}
	
\subsection*{Elastic dipolar interactions between model cells induce network formation} 
We expect the multicellular structures resulting from the dipolar cell-cell interactions to depend on three crucial nondimensional combinations of model parameters: the ratio of a characteristic elastic interaction energy $\mathcal{E}_{c}$, to noise  \textendash \hspace{.05cm} denoted by $A = \mathcal{E}_{c}/k_{B}T_{\mathrm{eff}}$ \textendash \hspace{.05cm} the effective elastic interaction parameter; the number of cells $N$, equivalently expressed as a cell density or packing fraction, $\phi = \frac{\pi N  \sigma^{2}}{4 L^{2}}$; and Poisson's ratio, $\nu$, which determines the favorable configurations (both position and orientation) of a pair of dipoles. To show the types of multicellular structures that result from our model elastic interactions, we perform Brownian dynamics simulations (detailed in \emph{Methods}) to generate representative snapshots at slices of this $A-\phi$ parameter space for two values of $\nu$; $0.5$ and $0.1$ shown in Figs. \ref{fig:Figure2} and SI Supp. Fig. 4, respectively. 
As packing fraction is increased, networks form more readily. As the effective elastic interaction is increased, cells form into networks characterized by chains, junctions, and rings. This can be thought of  naturally as a competition between entropy and energy. At low packing fractions or effective elastic interaction, cells are either in a gas--like state or form local chain segments with many open ends which have high entropy. As packing fraction or effective elastic interaction increases, cells relinquish translational and rotational freedom for more energetically favorable states such as longer chains, junctions, or rings. This is consistent with the cell density dependent  morphologies seen in images from \textit{in vitro} hydrogel experiments, shown in Fig. \ref{fig:Figure2}.  

We choose two representative values of $\nu$ in our model simulations because their corresponding strain plots are qualitatively different \cite{Bischofs2004} as seen in Fig. \ref{fig:MasterPlot2}e. Briefly, since contractile dipoles prefer to be on stretched regions of the substrate, the low (high) $\nu$ deformation patterns are expected to favor two (four) nearest neighbors. The different values of Poisson ratio could  correspond to synthetic hydrogel substrates and the fibrous extracellular matrix, respectively. While hydrogel substrates are nearly incompressible ($\nu = 0.5$), the ECM comprises of networks of fibers which permit remodeling and poroelastic flows leading to reduced material compressibility (e.g., $\nu = 0.1$) at long time scales\cite{Javanmardi2021-sy}. 



\subsection*{Substrate deformation-mediated interactions strongly enhance percolation in model networks}	
To characterize the extent of multi-cellular network formation, we consider the percolation order parameter which measures the ability of a connected network to span the available space. 
 Percolation is defined as the probability that, for a final realization of the network, there exists a continuous path through the network that spans the length of the simulation box. To compute percolation probability, we first identify connected clusters of cells, a process detailed in SI section E. A specific network configuration is considered to be percolating if  any two cells within the same cluster are separated by a Euclidean distance greater than or equal to the simulation box size. This calculation is done at the final time step of forty simulations per data point for dipole simulations, and ten simulations per data point for sticky disks shown in Fig.~\ref{fig:Figure3}. The average values and corresponding errors are then plotted as a function of packing fraction $\phi$ in Fig.~\ref{fig:Figure3}a and of the effective elastic interaction parameter $A$ in Fig.~\ref{fig:Figure3}b.

To contrast with the dipoles that mutually align through long-range and anisotropic interactions, we consider a control system of ``diffusing sticky disks''. These agents just diffuse without any long-range interactions and cease movement upon contact with another agent. We find percolating networks for both interacting elastic dipoles and diffusing sticky disks. However, Fig.~\ref{fig:Figure3}a shows model cells which interact as dipoles at long-range require far fewer cells to percolate than their sticky disk counterparts given that the elastic interaction strength is sufficiently greater than noise as shown in Fig.~\ref{fig:Figure3}b ($A\gtrapprox1$ in the case shown where $N=300$). This is because the anisotropic nature of the dipolar interactions promotes end-to-end alignment of cells, creating elongated structures like chains which can self-assemble into space-spanning networks. We therefore show that network formation requires fewer cells  when cells can sense, move and align in response to the substrate deformations created by other cells. Thus, networks guided by mechanical interactions are more cost efficient than when cells move or spread randomly, forming adhesive contacts upon finding their neighbors. 


\begin{figure*}[hbt!]  
	\centering
	\includegraphics[width=1\linewidth]{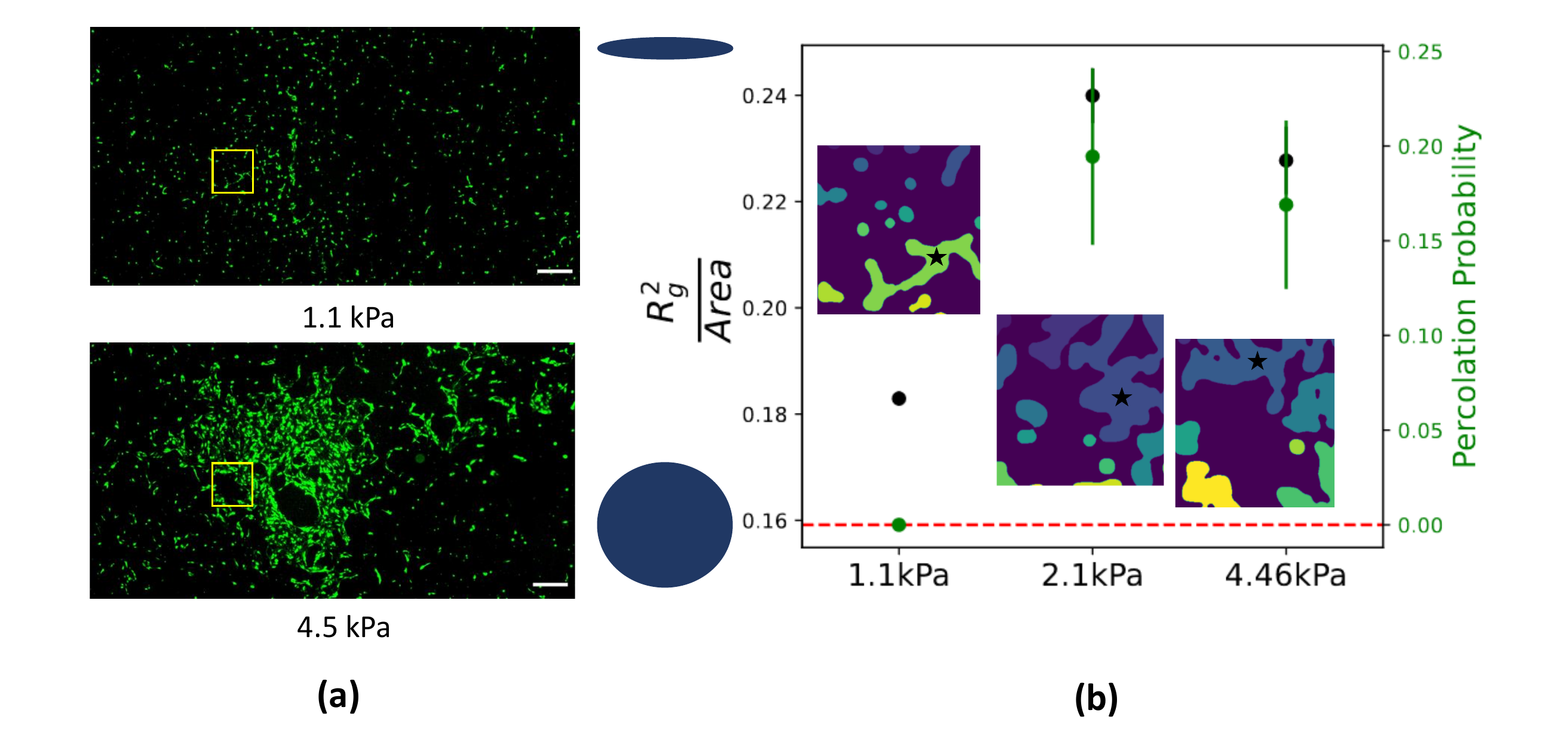}
	\caption{
    Analyis of connected clusters of endothelial cells cultured on hydrogel substrates reveals optimal stiffness for cell clustering.
    (a) Experimental images of HUVECs 16 hrs post seeding on polyacrylamide substrates of various stiffness: 1.1kPa (top) and 4.5kPa (bottom). Scale bars = 500\text{$\mu$}m. (b) Calculated metrics of percolation probability (green) and  size of largest cluster (black) are shown vs. substrate stiffness. Cells percolate, i.e. form long range connections, only on substrates of stiffness $\geq$ 2.1kPa. Cluster size is measured as the square of its radius of gyration ($R_{g}^{2}$) ) normalized by its area. Small values of this metric indicate a compact, isotropic shape with the lower bound (red dashed line) being a solid circular disk, while larger values indicate a branched or elongated  shape, \emph{e.g.} an ellipse. 
    Insets show local clusters labeled by color, with stars indicating the largest local cluster used in the $R_{g}^{2}$ analysis. Cell clusters are most spread out 
    at an intermediate substrate stiffness - 2.1kPa.}
	\label{fig:Rg2} 
	\end{figure*}

Much work has been done on characterizing the connectivity percolation transition on various lattice configurations  \cite{stauffer1979scaling}. The critical packing fraction can be widely different depending on the lattice geometry, and whether the space-spanning clusters comprise sites or bonds \cite{Yonezawa1989,Kirkpatrick1973}. The critical packing fraction for site percolation is known to be $\phi_{C} = 0.5$ for an infinitely large triangular lattice \cite{Sykes1963}. In approximate agreement with this, we find that the critical packing fraction for diffusive sticky disks for the current finite system size $L$ is $\phi_{C} \approx 0.6$.
For the dipolar particles, anisotropic interactions shift the percolation transition to $\phi_{C} \approx 0.2$, similar to those seen in dipolar colloidal assemblies at low reduced temperature \cite{schmidle_hall_velev_klapp_2011}.

Our observed packing fractions for transition to percolation are specific to the simulation system size, $L$, and differ from the actual  critical packing fraction due to finite size effects. How prominent these effects will be depends on the fractal dimension, which  provides a measure of how these structures scale with size. Since area scales like $L^{2}$, but number of particles scales like $L^{d_{f}}$, where $d_{f}$ is the fractal dimension, $\phi_{C} \propto L^{d_{f}-2}$. Therefore, there exists a regime in which $\phi_{C}$ will decrease with increasing $L$, as shown by simulations with bigger box sizes shown in SI section F. A more in depth analysis of the fractal dimension of these systems is presented later.



\begin{figure*}[hbt!]  
	\centering
	\includegraphics[width=1\linewidth]{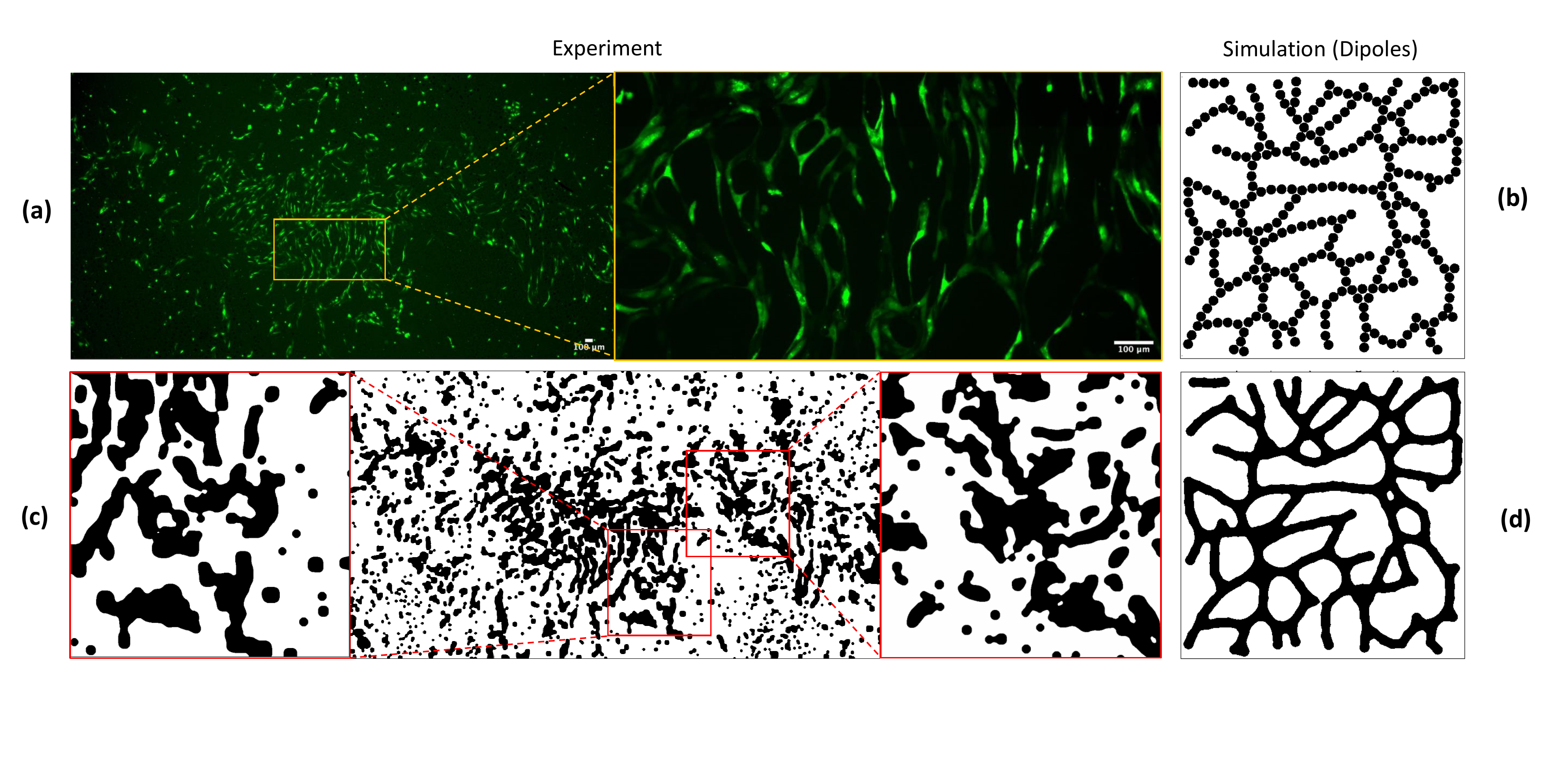}
	\caption{Model cell dipole networks exhibit fractal dimensions similar to cell culture experiments. (a) Fluorescence microscopy image of HUVECs seeded on polyacrylamide substrate of stiffness $E = 2.1\text{kPa}$. Inset shows a zoomed in region of highly branched and ringed structures qualitatively similar to the predominant morphologies of simulated dipole networks. (b) Simulated dipolar particles in a region of parameter space in which space-spanning networks form. (c) Processed image of (a) where the processing scheme is described in depth in Appendix D. Insets show regions of processed experimental images wherein the local area fraction is similar to those in simulation. (d) Processed image of (b) where the processing scheme is described in depth in Supplementary Information.}
	\label{fig:Figure3.5} 
	\end{figure*}

\subsection*{Elastic interactions are optimal at intermediate substrate stiffness}

The percolation dependence can be mapped from the effective elastic interaction parameter, $A$, to substrate stiffness, $E$, by using an experimentally motivated dependence of cell forces on substrate stiffness, $P= P_{0} E /(E+E^{*})$. Here, $P_{0}$ corresponds to the maximum traction generated by a well-spread cell on a very stiff substrate, and $E^{*}$ is the substrate stiffness at which the cell force saturates \cite{Ladoux2017}. The resulting elastic interaction parameter, $A$, is weak on soft substrates where cell forces are low and on stiffer substrates, where the deformations are low, as detailed in SI section G.  
This mapping from effective elastic interaction to substrate stiffness (SI Supp. Fig. 4) results in a peak in the percolation curve  Fig.~\ref{fig:Figure3}c over an interval of substrate stiffness at the optimal stiffness $E^{\ast}$, which depends on both cell density and effective temperature representing noisy cell movements. Higher effective temperature and lower cell density reduce both peak height and width. This result is consistent with experiments on EC cultures (Fig.~\ref{fig:MasterPlot2}a, b) which show that percolating networks form only in a certain range of substrate stiffness.  The optimal stiffness is expected to be specific to cell type and matrix properties such as ligand density. Thus, a shift in $E^{\ast}$ in our model may explain the findings of Ref.~\cite{Califano2008} that EC networks form on softer substrates at higher density of collagen ligands, while they prefer stiffer substrates on substrates with lower collagen density. It is plausible that when fewer ligands are available to adhere to, the cells require stiffer substrates to spread more and reach maximum contractility \cite{ENGLER2004617}.

\subsection*{Analysis of experimental cell culture images suggests substrate stiffness dependent cell network formation }

To test the prediction of an intermediate substrate stiffness at which cell network formation is optimal, we performed 2D cell culture experiments on elastic substrates. Human umbilical vascular endothelial cells (HUVECs) were seeded on  fibronectin-coated polyacrylamide substrates of varying stiffness. The substrate preparation protocol, described in \textit{Methods}, and stiffness characterization of these substrates follow standard precedents \cite{Tse2010}. After sixteen hours elapse, fluorescence images were taken (Fig.~\ref{fig:Rg2}a). Over these timescales, cells can spread and form contacts, but do not typically divide. We divided the large field of view in each experimental image into 73 sub-boxes for more statistics, and analyze each sub-box for cell cluster formation using ImageJ \cite{Schneider2012} (see \textit{Methods}). We then checked for connectivity percolation of all the sub-boxes(Fig.~\ref{fig:Rg2}b - green). We find that the percolation probability is insignificant for the 1.1 kPa soft substrate, while that on substrates of stiffness 2.1 kPa and 4.5 kPa are nearly identical. This is not surprising because cells spread less on the 1.1 kPa substrate, and result in significantly less fractional area coverage compared with the stiffer substrates. Thus, a global measure like percolation may not reveal the tendencies of cells to cluster locally at lower density.

In order to obtain a measure of how spread out each cell cluster is, we calculate a ``shape parameter'', defined as $\frac{R_{g}^{2}}{Area} \equiv \frac{1}{N^{2}} \sum_{k=1}^{N} (\mathbf{r}_{k}-\mathbf{r}_{COM})^{2}$, for the largest cluster in each sub-box. Here, $R_{g}$ represents the radius of gyration of the cluster, which is defined about its center-of-mass $\mathbf{r}_{COM}$, and $N$ is the number of occupied pixels in each cluster.  The normalization by cluster area ensures we control for cluster size variations between different experiments. Lower values of this shape parameter correspond to isotropic shapes, the lower bound being $\frac{1}{2\pi}$ for a solid  circular disk(Fig.~\ref{fig:Rg2}b - red dashed line). Conversely, a higher shape parameter corresponds to more anisotropic shapes such as high aspect ratio ellipses. This is a suitable proxy measure of global connectedness via local elongation. We find that this metric gives us the same result as percolation for the very soft case (1 kPa), where cells are by and large isolated, but the intermediate stiffness (2.1 kPa) exhibits a statistically significant greater shape parameter value than 4.5 kPa. This non-monotonicity lends credence to the prediction of our model that network formation is induced via substrate mediated elastic interactions which are optimal within some interval of substrate stiffness values 




We now turn  to evaluating the morphological similarity of our simulated dipoles and our experimental cell culture by calculating its fractal dimension. In order to compare our experimental cell clusters with our simulated dipole networks, we skeletonize the binary images of experiments and simulations (Fig.~\ref{fig:Figure3.5}c,d) and obtain average branch lengths. The ratio of the average branch length in experiment to that in the simulation is multiplied by length and width of the simulation box size to obtain a sub-box length and width, respectively. The full experimental image is then parsed into corresponding sub-boxes. We then check all these regions and keep only those sub-boxes, whose local area coverage is between 30-40\%, so as to omit outlier regions where there are too few or too many cells. These sub-boxes, three characteristic simulation final states for $\nu=0.1$ and $\nu=0.5$, and the control case of purely diffusive sticky disks at percolation are analyzed via ImageJ's fractal box count function. 

For the ``sticky disks'', we find a fractal dimension of $d_{f}=1.81$, whereas for the dipoles we find fractal dimensions of $d_{f} = 1.698 \pm .004$ and $d_{f} = 1.711 \pm .003$ for  $\nu = 0.1$ and $\nu = 0.5$, respectively (Fig.~\ref{fig:Figure3.5}b,d). We find a similar fractal dimension for our experimental HUVEC culture on a substrate of stiffness $E = 2.1\text{kPa}$, $d_{f} = 1.712 \pm .003$ (Fig.~\ref{fig:Figure3.5}a,c). Interestingly, simulated networks on substrates of $\nu=0.1$ and $\nu=0.5$ are statistically distinguishable, with the experimental fractal dimension showing excellent agreement with the $\nu =0.5$ simulated dipole case . This is in accordance with the approximately incompressible nature of hygrogel substrates. The proximity of the fractal dimensions of the simulated dipoles to that of experimental cell networks, in relation to the sticky disks, indicates that cells utilize a more complex strategy to self-assemble than simply random movement followed by cell-cell adhesion. The elastic dipolar interactions are thus a plausible strategy allowing the self-assembly of biologically desirable, space-spanning and cost effective networks.


\begin{figure*}[hbt!]  
	\centering
	\includegraphics[width=\linewidth]{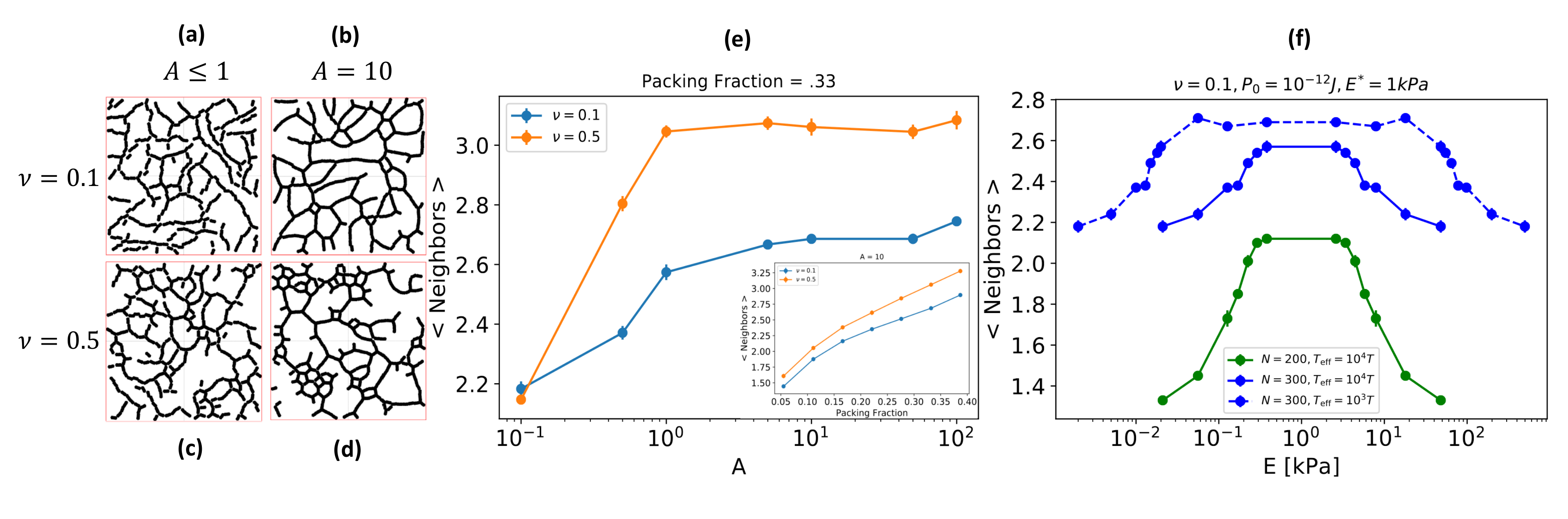}
	\caption{Neighbor counts reveal relative prevalance of various morphological structures in networks formed by elastic dipolar interactions.(a)-(d) Simulation snapshots of cell assemblies at the shoulder of the percolation transition (left) and well above the percolation transition (right).
	(e) Number of neighbors as a function of A when $\phi = .33 (N=300)$ for $\nu = 0.1$ - blue - and $\nu = 0.5$ - orange where a neighbor in this context is defined as a cell $\alpha$ whose center is within one and a half cell diameters away from cell $\beta$ ($|r_{\alpha\beta}| \le 1.5 \sigma$). While the number of neighbors is relatively insensitive to A, there is a marked difference between the two values of Poisson's ratio. Across $A$ space, cells on substrates of higher $\nu$ values accumulate more neighbors than the lower $\nu$ cases. Inset shows number of neighbors as a function of packing fraction for $A = 10$. Cells on higher $\nu$ value substrates have more neighbors than the low $\nu$ case regardless of packing fraction. (f) Number of neighbors as a function of substrate stiffness. Optimal stiffness is assumed to be 1 kPa. $N = 200(\phi \approx .22)$ exhibits an average neighbor count of 1-2 indicating the prominence of short chains. $N = 300(\phi \approx .33)$ case shows average neighbor counts of 2-3 indicating an abundance of chains, rings, and junctions. The peak in neighbor count over stiffness is taller and wider for lower effective temperature  and higher cell density.}
	\label{fig:Figure4} 
	\end{figure*}

\subsection*{Model network morphological features depend on substrate compressibility given by Poisson's ratio}

While percolation is by nature a global quantity describing the whole network, we now employ more local metrics to classify the topology of our networks. Figs. \ref{fig:Figure4}a-d show characteristic networks of $N=300$ ($\phi=0.33$) cells for $\nu=0.1$ (top) and $\nu=0.5$ (bottom) when the system is well past the percolation transition ($A=10$, right), and at the shoulder of the transition ($A \leq 1$, left). The particles in these snapshots have been given artificially elongated bodies along their dipole axis to emphasize the backbone of the network and aid the image analysis process, detailed in SI section E. The relative number of the different topological features of these networks, \emph{e.g.} open ends, junctions, and rings, will determine the average number of neighbors (or coordination number, $z$) of each cell dipole.


Fig. \ref{fig:Figure4}e shows that the average number of neighbors increases with effective elastic interaction $A$ (for $N = 300$ fixed) and cell number density (for $A = 10$ fixed), for both $\nu = 0.1$ and $\nu = 0.5$ that saturates in $A$. This quantity is calculated for the final simulation configuration of three networks per value of Poisson’s ratio. The saturating neighbor count for each Poisson's ratio is reached for percolating networks, and corresponds to the disparate topological features characteristic of these two cases. The higher $\nu = 0.5$ (incompressible) substrate case shows a higher saturating neighbor ($z>3$), which indicates the preeminent structures inherent to this network are junctions and tighter rings (with up to $4$ neighbors), consistent with the characteristic simulation snapshots in  Fig.~\ref{fig:Figure4}d.  The saturating neighbor count for low  $\nu = 0.1$ (more compressible) substrates is lower ($2<z<3$. This suggests that these networks exhibit long chains as well as more interconnected structures like junctions and rings, consistent with the characteristic simulation snapshots shown in Fig.~\ref{fig:Figure4}b. This trend is seen over a wide range of packing fractions as shown by the inset in Fig. \ref{fig:Figure4}e. The qualitative differences between the two types of networks ultimately arise from the different orientational dependencies of the deformation induced by a dipole, as shown in Fig.~\ref{fig:MasterPlot2}e, with a transition expected at $\nu=0.3$ \cite{Bischofs2005}.  We note that these results are for a relatively dilute regime ($\phi =0.33$), whereas in the limit of complete packing,  neighbor count would saturate to a maximum possible value of $6$. 

Interestingly, the networks on the lower Poisson ratio substrates exhibit a saturating neighbor count ($z \simeq 2.6$), that is very close to that for the predicted rigidity percolation threshold for elastic fiber networks ($z_c = 2.67$) \cite{Huisman2011, Zaccone2013elastic}.  This may be attributed to the self-assembled linear chains that mimic semiflexible polymers \cite{Broedersz2014modeling}, with bending rigidity of a ``polymer'' of disks, set by the dipolar interaction strength, $A$. This implies that although we do not measure the rigidity of networks in simulation, the connectivity percolation is closely related to it and predicts the onset of rigidity percolation threshold as well. Such a transition from isolated, fluid-like, motile cells to a mechanically rigid state has been shown to be biologically important for epithelial cells during development and disease \cite{petridou2021rigidity} , and may also be relevant to network-forming endothelial cells.  

Similarly to percolation, Fig. \ref{fig:Figure4}f shows that neighbor counts exhibit peaks over intervals of substrate stiffness centered around the characteristic substrate stiffness (chosen to be $E^{\ast} = 1$ kPa) and can be narrowed and decreased by increasing effective temperature and decreasing packing fraction. This result is consistent with Fig.~\ref{fig:MasterPlot2} where cells on substrates that are too soft or too stiff remain isolated, and have fewer neighbors than those in network configurations that form at the optimal stiffness range.

\begin{figure*}[hbt]  
	\centering
	\includegraphics[width=1 \linewidth]{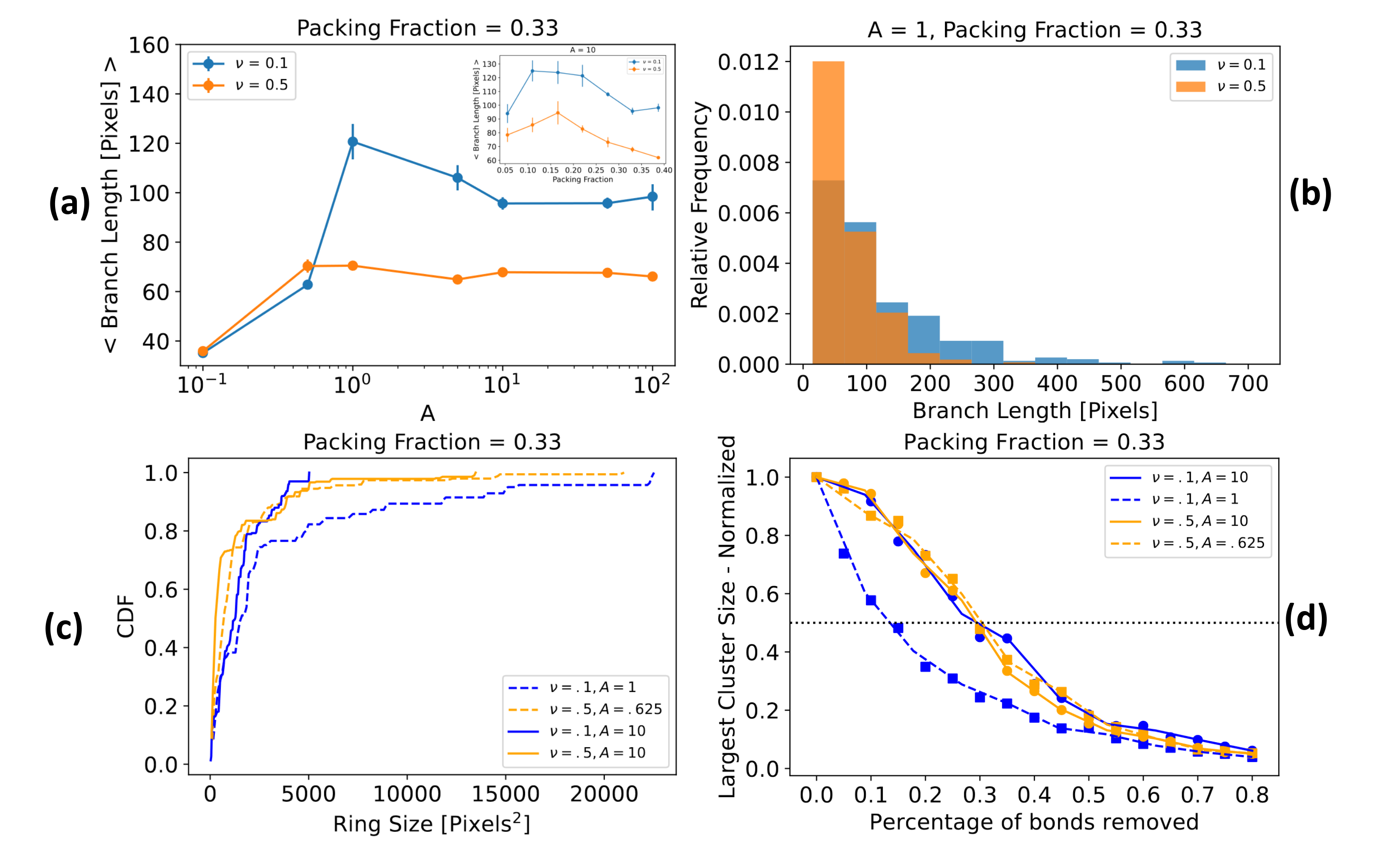}
	\caption{Substrate compressibility and rigidity affect  efficiency and resilience of model networks. (a) Average branch length as a function of the effective elastic interaction for $N = 300 (\phi \approx .33)$ cells. The lower $\nu$ case shows a greater sensitivity to $A$ indicating a greater aptitude for tunability than the high $\nu$ counterpart. The inset shows average branch length as a function of packing fraction when $A=10$. Both values of $\nu$ show similar behavior except at the highest point of packing fraction. At this packing fraction, the curves diverge as global configurations begin to become prevalent. For the low $\nu$ case, this will be long parallel strings whereas the high $\nu$ case will form a single cluster of 4-rings. (b) Normalized branch length histogram for $A = 1$ and $\phi = .33$. The networks on substrates of high $\nu$ are sharply peaked around the smallest branch lengths while the networks at low $\nu$ exhibit a broader, longer-tailed distribution. (c) Cumulative distribution of ring area for $N = 300 (\phi \approx .33)$ cells shown both for networks at the shoulder of the percolation transition and networks well beyond the transition. Networks at high $\nu$ contain smaller rings than the networks at low $\nu$. Irreversible networks show more smaller rings as noise is not great enough to jostle these compact structures apart to favor more stringy morphologies. (d) Largest cluster size as a function of the fraction of network branch segments removed - a measure of a networks ability to to maintain functionality after being damaged \cite{Javanmardi2021-sy}. Networks at the shoulder of the percolation transition exhibit less robustness than those well above the percolation transition 
	for the $\nu=0.1$ case. In the $\nu=0.5$ case, however, networks retain their robustness even at the shoulder of the percolation transition. As this robustness metric saturates at a value of $A$ dependent on the compressibility of the substrate, we hypothesize cells interacting in the way that we have estimated will tend to exert only a certain amount of force,  enough to build a resilient network and no more.} 
	\label{fig:Figure5} 
	\end{figure*}
 
\subsection*{Diverse morphological features offer distinct advantages in network assembly and transport function} 

We now analyze the two predominant network morphological constituents - chains and rings - over the parameter space and relate the results to transport function. 
Fig. \ref{fig:Figure5}a shows average branch length for $N=300$ ($\phi = 0.33$) cells, where each data point represents final simulation configurations of three networks per value of Poisson’s ratio, as a function of effective elastic interaction ($A$). 
The average branch length for the higher $\nu$ case remains constant and low at just over 60 pixels (about two cell lengths), over our range of elastic interaction strengths. The lower $\nu$ case exhibits a peak in average branch length at the percolation threshold ($A=1$) before decreasing and saturating at high $A$ values. The distribution of branch lengths (Fig.~\ref{fig:Figure5}b) shows that while $\nu=0.5$ is sharply peaked near the smallest branch sizes ($\sim$ 30 pixels) , $\nu=0.1$ exhibits branches greater than 600 pixels and shows a greater relative count in the 100-400 pixel range than the higher Poisson's ratio counterpart. 

These results suggest that at higher values of $\nu$, network morphology is more resilient to noise, and the branch lengths are not as easily tunable, 
The greater variability in branch lengths leads to longer branches in the lower $\nu=0.1$ case, which then requires  (for $A \geq 5$) fewer cells to percolate than at $\nu=0.5$. This is  seen by the difference of the curves at the shoulder of the percolation transition in Figs.~\ref{fig:Figure3}a and Supp. Fig. 10. The greater resilience of the network at higher substrate $\nu$ leads to percolation at smaller $A$ than its low $\nu$ counterpart (Figs.~\ref{fig:Figure3}b and Supp. Fig. 10). In SI section I, we construct a detailed map of the percolation transition in the $A-\phi$ parameter space, to show how $\nu=0.1$ requires fewer cells to percolate for a range of $A$ values, while $\nu=0.5$ can percolate at lower values of $A$ (Supp. Fig. 10).  This suggests that the two regimes of substrate compressibility optimize two different measures of cost of network building: one,  the number of cells, and the other, the strength of cell contractility.

Fig.~\ref{fig:Figure5}c shows a cumulative distribution of ring area for our networks at two crucial regions in our parameter space -- those at which the networks are well above the percolation transition (solid lines), or just above it (dashed lines). Similar to the branch length distribution, the networks at higher $\nu$ form many small rings and few large rings, while the lower $\nu$ case shows a broader distribution of ring sizes.  The tendency of the $\nu=0.5$ configurations to form more and smaller rings leads to a marginally less efficient area coverage than the low $\nu$ case, which forms longer branches resulting in less frequent small rings (Supp. Fig. 11).  These results are also consistent with the fractal dimensions we obtained earlier, with $d_{f}$ being slightly higher for the $\nu=0.5$ than the $0.1$ cases, indicating more compact structures for the former.

\subsection*{Network robustness correlates with ring formation}
To examine the robustness of our model networks to damage, a biologically significant property, we measure the largest remaining cluster size as a function of the fraction of network bonds removed \cite{Papadopoulos2018-xi} (Fig. \ref{fig:Figure5}d). 
We find that whether well above or just at the percolation threshold, the networks at higher $\nu$ retain cluster size well as bonds are removed. Networks at lower $\nu$ well above the percolation transition lose largest cluster size at the same rate as their higher $\nu$ counterpart. At the shoulder of percolation, however, networks at low $\nu$ lose largest cluster size and fall apart much more rapidly than any of the other networks. This is the same parameter regime at which networks at low $\nu$ exhibit a peak in branch length. By forming long branches, ring structure formation is sacrificed. Thus, we find that the prime factor for robust networks is the tendency to form rings which provide degeneracy to paths between any two nodes in the network - a result consistent with network structure optimization models \cite{Ronellenfitsch2016}.  
In summary, at lower $\nu$, networks tend to form longer and more broadly distributed branches which promote efficiency with respect to the filling and spanning of space at the cost of being susceptible to damage, while at higher $\nu$, networks are predominantly composed of small rings, which provide robustness to the networks at the cost of transport efficiency.

\section*{Discussion} Our model generates testable predictions for the dependence of cell network morphology on substrate mechanical properties. By performing and analyzing experiments on ECs cultured on hydrogels of varying stiffness, we show that network formation is indeed optimized at an intermediate stiffness. Although many experiments show that EC network formation or capillary sprouting require softer matrices (see Ref.~\cite{Crosby2019} and references therein), these findings can show different trends at different stiffness regimes, as shown in  Fig.~\ref{fig:MasterPlot2}\cite{Mason2013, Berger2017}. We suggest that this maybe because cells adapt their traction forces to substrate stiffness, and therefore the expected optimal stiffness for network formation should be where cells attain their maximal contractility. This optimal stiffness may be dependent on cell type and matrix mechanochemistry \cite{Califano2008}. 

Our modeling thus relates network structure to cell contractility, and the predictions can be further checked in cell culture experiments on substrates of varying stiffness and Poisson's ratio \cite{Javanmardi2021-sy}, that combine traction force measurement with quantification of network morphology. The presence of substrate deformation-mediated interactions can also be directly investigated in a two-cell setup on a micropatterened substrate which allows one to observe reorientations of one cell in response to the other, similar to strategies used to examine pairwise interactions during cell motility \cite{Brckner2021} and cardiomyocyte synchronization \cite{Nitsan2016}.


Further, cells may persistently migrate, in addition to the stochastic movements assumed in the present model. Our prior work suggests that cells form stable network structures rapidly at lower migration speeds \cite{BoseNoerr2022}. At high persistent migration speeds, the networks dissolve and the dipoles self-organize instead into motile chains. This suggests that an optimum cell migration speed is favorable for network formation, which cells may achieve through self-regulation of their motility through interaction with their neighbors, such as contact-inhibition of locomotion.

A crucial modeling challenge for vasculogenesis, and other instances of cell network formation in biology, is that multiple factors ranging from cell differentiation to chemotactic cues could be involved \textit{in vivo}. Modeling approaches based on different hypotheses can all lead to network pattern formation \cite{Scianna2013}. Here, by combining experiments on hydrogels of varying stiffness and a physical model based on mechanical interactions alone, we aim to isolate the different factors involved. While we focus on endothelial cell networks as a model system, our predictions are generally applicable to other contractile cell types that self-organize into networks such as fibroblasts \cite{Doha2022}, neurons or smooth muscle cells (Table S1), as well as to synthetic particles with electric or magnetic dipolar interactions, that are of interest in materials science.
In summary, our work provides proof-of-concept that substrate-mediated elastic interactions is a physical strategy that biological cells may employ to direct their self-organization into efficiently space-spanning, multicellular networks.

\section*{Methods}

\appendix


\section{Model details}

We model the ubiquitous traction force pattern of a polarized cell as a single, anisotropic force dipole. The dipole magnitude is the cell force times the distance along the long axis of the cell, $P = F a$. Since the contractile cytoskeletal machinery (e.g. actomyosin stress fibers) of the cell is typically aligned along this axis, this is also usually the principal direction of stress exerted by the cell and is henceforth called the ``dipole axis''. Such a force dipole induces a strain in the substrate, which is modeled as an infinitely thick, linear, isotropic elastic medium.

By considering two dipoles $\mathbf{P}^{\alpha}$ and $\mathbf{P}^{\beta}$, we  show in SI section A  that the work done by a dipole $\beta$ in deforming the elastic medium in the presence of the strain created by the other dipole $\alpha$, is given by \cite{Bischofs2004}: $W^{\alpha \beta} =  P^{\beta}_{il} u^{\alpha}_{il}({\mathbf r}^{\beta})$, where the strain can be written in terms of $\mathbf{P}^{\alpha}$ and second derivatives of an elastic Green's function as $u^{\alpha}_{il}(\mathbf{r}^{\beta})=\partial_{l}\partial_{k}G_{ij}(\mathbf{r}^{\beta}-\mathbf{r}^{\alpha})P^{\alpha}_{jk}$. This minimal coupling between dipolar stress and medium strain represents the mechanical interaction energy between dipoles.
Typical substrate strains are shown in Fig.~\ref{fig:MasterPlot2}(e) where the blue (red) coloring represents expanded (compressed) regions. A second or test dipole present in these regions would tend to align its contractile axis along the principal stretch direction of the substrate. In the expanded (blue) regions, the test dipole is aligned with and attracted towards the central dipole, whereas in the compressed (red) regions, a test dipole is  aligned orthogonal to and repelled away from the central dipole. The orientational dependence of the strain field is changed by  the Poisson's ratio or compressibility of the substrate \cite{Bischofs2003}. 

Our computational ``many-cell'' model considers cells as discrete agents ($N$ agents in a $L \times L$ box with periodic boundary conditions) which move and orient randomly, but that also interact with one another through long-range elastic interactions via a force dipole strain field coupling and a short-range repulsive spring. Fig.~\ref{fig:MasterPlot2}(f) shows our simulation setup and the main ingredients of the model. We ignore details of the cell shape and subcellular structures in this minimal model and instead consider the cells as disk-shaped agents endowed with contractile,elastic dipoles.  This simplifying assumption implies that we do not consider changes in the shape and size  of individual cells that occur as a result of cell-substrate feedback when substrate stiffness is varied, but instead focus on the multicellular structures at longer length scales.


We now consider the translational and orientational dynamics of a collection of model cells. These interact with each other through short range, steric and long-range, substrate-mediated, elastic interactions, and undergo diffusive motion.
The overdamped Langevin dynamics governing the position of a cell labeled $\alpha$ is 
    \begin{equation}
    {\frac{d\mathbf{r}_{\alpha}}{dt}} =  -\mu_{T}\sum_{\beta} \frac{\partial W_{\alpha \beta}}{\partial {\bf r}_{\alpha}} + \sqrt{2D_{\mathrm{T}}}\:\: \bm{\eta}_{\alpha,\mathrm{T}}(t)
    \label{model_eq}
    \end{equation}
    where $D_{\mathrm{T}}$ is the effective translational diffusivity quantifying the random motion of an isolated moving cell, with $\bm{\eta_{\mathrm{T}}}$ as a random white noise term whose components satisfy 
    $\langle \eta_{i,\mathrm{T}}(t) \eta_{j,\mathrm{T}}(t') \rangle = \delta(t-t') \delta_{ij}$. While cell movements in principle are persistent, the delta-correlated white noise assumption is valid if we consider cell displacements over time scales that are longer than this persistence time.
 The mobility  $\mu_{T}$ in Eq.~\ref{model_eq} is inversely related to the effective friction from the medium that the moving cell experiences at its adhesive contacts with the substrate. Similarly, the orientational dynamics of the cell denoted by $\alpha$ is given by 
   \begin{equation}
    {\frac{d\hat{\mathbf{n}}_{\alpha}}{dt}} =  -\mu_{R} \sum_{\beta}\hat{\mathbf{n}}_{\alpha}\times\frac{\partial W_{\alpha \beta}}{\partial \hat{\mathbf{n}}_{\alpha}} + \sqrt{2D_{\mathrm{R}}}\:\: {\bm{\eta}}_{\alpha,\mathrm{R}}(t)
    \end{equation}
 where $\hat{\bm{n}}_{\alpha}$ is the unit vector along the dipole axis of the cell $\alpha$ and $D_{\mathrm{R}}$ is the effective rotational diffusivity quantifying the random reorientations of an isolated moving cell.  Cells encounter various forms of internal stochastic effects including internal cytoskeletal rearrangements producing membrane morphological fluctuations, substrate surface binding fluctuations and fluctuations in myosin motor forces, which are all absorbed into a coarse-grained effective temperature, $T_{\mathrm{eff}}$, in our model. Single cell and cell cluster experiments have shown this effective temperature to be on the order of $10^{-15}-10^{-14}$ J \cite{Beysens2000}. Though the rotational and translational diffusion are in principle independent, we will here assume them to correspond to the same underlying processes and therefore the same effective temperature, $k_{B}T_{\mathrm{eff}} = D_{T}/\mu_{T} = D_{R}/\mu_{R} $. We also show some exceptions to this assumption in the SI section L, which all robustly form networks.

    
The pairwise cell-cell interaction potential $W_{\alpha \beta}$ between cells labeled $\alpha$ and $\beta$ consists of the long-range elastic interaction arising through their mutual deformation of the substrate (SI section A), and a short-range steric interaction between two cells in contact, and is given by,
\begin{eqnarray}
    W_{\alpha \beta} &=& 
        {\frac{1}{2}}k {(\sigma - r_{\alpha \beta})^2}, \:\:\: {\mathrm{when}} \:\:\: 0\leq r_{\alpha \beta}\leq \sigma \:\:\: \:\:\:  \nonumber \\   
        &=&
        {\frac{P^{2}}{E}} {\frac{f(\nu,\theta_{\alpha},\theta_{\beta})}{r_{\alpha \beta}^{3}}}, \:\: \:\:\: {\mathrm{when}} \:\:\: r_{\alpha \beta} > \sigma ,
    \label{potential_eq}
\end{eqnarray}
where $f$ is a function of Poisson's ratio - shown in SI section A, $\theta_{\alpha}$, and $\theta_{\beta}$ where $\cos \theta_{\alpha} = \bm{\hat{n}}_{\alpha}\cdot\hat{\bm{r}}_{\alpha \beta}$ and $\cos \theta_{\beta} = \bm{\hat{n}}_{\beta}\cdot\hat{\bm{r}}_{\alpha \beta}$ are the orientations of cell $\alpha$ and cell $\beta$ with respect to their separation vector, $\mathbf{r}_{\alpha \beta} = \mathbf{r}_{\beta} -  \mathbf{r}_{\alpha}$ connecting the centers of the two model cell dipoles, respectively.  Note that while the elastic potential is in principle long-range, it decays strongly as a $1/r^{3}$ power law, we cut this pairwise interaction off at $r_{\alpha \beta} > 3 \sigma$ in our simulations, since the substrate strain induced by one cell is unlikely to be detected by a cell few cell lengths away \cite{Tang11}.


   
The above equations are non-dimensionalized by a suitable choice of length, time and energy scales.  By choosing the length scale to be the cell diameter $\sigma$, the time scale to be an elastic time, $t_{c} =\frac{16E\sigma^{5}}{P^{2}\mu_{\mathrm{T}}}$, and a characteristic elastic interaction as the energy scale, $\mathcal{E}_{c} =\frac{P^{2}}{16E\sigma^{3}}$ ,  the dynamical equations reduce to (Appendix B),
\begin{equation}
    {\frac{d\mathbf{r^{*}_{\alpha}}}{dt^{*}}} =  -\sum_{\beta}\frac{\partial W^{*}_{\alpha \beta}}{\partial {\bf r^{*}_{\alpha}}} + \sqrt{\frac{2}{A}}\:\: \bm{\eta^{*}}_{\alpha,\mathrm{T}}(t^{*}),
    \end{equation}
for the translational motion,while the rotational equation of motion can be written as, 
    \begin{equation}
    {\frac{d\hat{\bm{n}}_{\alpha}}{dt^{*}}} =  -\sum_{\beta}\hat{\mathbf{n}}_{\alpha}\times\frac{\partial W^{*}_{\alpha \beta}}{\partial {\hat{\mathbf{n}}_{\alpha}}} + \sqrt{\frac{2}{A}}\:\: {\bm{\eta^{*}}_{\alpha,\mathrm{R}}}(t^{*}),
    \end{equation}
where the starred variables indicate nondimensionalized quantities, and we have assumed $\mu_{\mathrm{R}}\sigma^{2}=\mu_{\mathrm{T}}$ and $D_{\mathrm{R}}\sigma^{2}=D_{\mathrm{T}}$, although the latter is not required for a system that is out of equilibrium.  The nondimensionalized  pairwise interaction potential in Eq.~\ref{potential_eq} is here given by     $W^{*}_{\alpha \beta} = \frac{1}{2}k^{*}(1-r^{*})^{2}\Theta({1-r^{*}}) -\frac{16 f}{r^{*3}}\Theta({r^{*}-1})$, where $k^{*}= k \sigma^{2}/\mathcal{E}_{c}$.  We introduce an effective elastic interaction parameter quantifying the  elastic interaction strength relative to that of intrinsic noise in the cell motion, 
\begin{equation}
   A =  \frac{P^{2} \mu_{T}}{16E\sigma^{3} D_{T}} = \frac{ \mathcal{E}_{c}}{k_{B}T_{\mathrm{eff}}},
   \label{eq_defA}
\end{equation}
where the noisy cell movements correspond to an effective temperature, $k_{B}T_{\mathrm{eff}} \equiv D_{T}/\mu_{T}$. This explicitly shows that $A$ is a measure of the characteristic elastic interaction energy scale relative to the magnitude of cell stochasticity described by an effective temperature. 

\section{Physiological estimates of parameter values}

In experiments, the value of the effective interaction parameter $A$ will depend on cell contractility, the stiffness of the elastic substrate, and the diffusivity that originates from the motility of single cells. Importantly, cells adapt their contractile forces to the stiffness of the underlying substrate. Measurements \cite{Ladoux2017} and models \cite{De2007} of the dependence of cell force on substrate stiffness suggest that the magnitude of the force dipole can be written as: $P(E) = P_{0}/( 1 + E/E^{\ast})$, where the characteristic substrate stifffness for a given cell at which the cell traction forces saturate to their maximal value is denoted by $E^{\ast}$. This dependence when inserted into the definition of the effective elastic interaction parameter, $A$, in Eq. ~\ref{eq_defA}, leads to $A$ being a peaked function of $E$.  Since stiffer substrates are harder to deform and cells on softer substrates don't generate enough traction, substrate deformations and therefore elastic interactions are maximal at an intermediate optimal stiffness value ($E= E^{\ast}$). 

\begin{table}[ht] 
\caption{Simulation parameters and their meaning.\label{tab1}}
\begin{center}
\begin{tabular}{lll}
\toprule
\textbf{Parameter}	& \textbf{Interpretation}	 & \textbf{Simulation values}\\
$A$ & 
 Elastic interaction : Noise & 0.1-100    \\
$k^{\ast}$ & steric interaction  & $1.6 \times 10^{3}$    \\
$\phi$ & Cell packing fraction & 0.05-0.5\\
$\sigma$ & Cell diameter & 1 \\
$L$ & Box size & 26.66 \\
\end{tabular}
\end{center}
\end{table}

To identify a plausible range for the values of $A$ consistent with cell culture experiments, we note that the typical values for the force dipole for contractile cells adhered to elastic substrates is $P_{0}= F\sigma\sim 10^{-12}-10^{-11}$ \si{J} \cite{Bischofs2005, Bischofs2006}. This corresponds  to measured traction forces of $ F\sim 10 - 100$ \si{nN} with a distance of $\sim 50$ $\mu$m separating the adhesion sites at which the forces act on the substrate \cite{Schwarz2002, Reinhart-King2003}, which is also the typical size of the cell along its long axis. For a typical substrate stiffness of $E\sim 1$ \si{kPa} characteristic of EC network formation \cite{Califano2008, Rudiger2020}, we therefore estimate an elastic dipole  energy of $\mathcal{E}_{c} = \frac{P^{2}}{16 E \sigma^{3}} = \frac{F^{2}}{16 E \sigma} \sim 10^{-15}$ \si{J}, similar to measured values for cell contractile energy stored in the elastic substrate \cite{Mandal2014}.  Since, adherent cells crawl by exerting forces at the focal adhesions at which forces are transmitted to the substrate, the net mobility that determines cell translation, $\mu_{T}$, can be estimated from the friction force at these adhesion sites. From the observation that the focal adhesions with surface area of $10$ $\mu$m$^{2}$ reorient with speeds of $\mu$m/min in the direction of an external, applied stress of kPa \cite{Riveline2001}, we can estimate the mobility coefficient (inverse of friction coefficient) to be $\mu_{T} \sim 0.1 \mu\mathrm{m/min} \cdot pN^{-1}$. The effective diffusivity characterizing single endothelial cell movements was measured to be $\sim 10$ $\mu \mathrm{m}^{2}/$min \cite{Stokes1991, Reinhart2008, Rudiger2020}. Together, these give an estimate for the effective temperature: $k_{B}T_{\mathrm{eff}} = D_{T}/\mu_{T} \sim 10^{-16}$ \si{J} $\sim 10^{4}$ $k_{B}T$. For substrate stiffness $E \sim E^{*}~ 1$ kPa, we thus estimate the ratio of elastic energy to noise to be $A = \mathcal{E}_{c}/k_{B}T_{\mathrm{eff}} \sim 10$.

In experiments, the substrate stiffness can be tuned over a wide range. In particular, Califano \textit{et al.} tested the formation of EC networks on substrates whose rigidity was varied from $100  \mathrm{Pa}-10 \mathrm{kPa}$ \cite{Califano2008}. This, in our estimate, corresponds to an interaction parameter $A \sim 1 - 100$, with $A=0.1$ corresponding to high noise or non-optimal values of substrate stiffness (too soft or too stiff).  Similarly, we can estimate the characteristic timescale as $t_{c} = \frac{\sigma^{2}}{\mathcal{E}_{c} \mu_\mathrm{T}} \sim 10^{2}$ min. This timescale of hours is consistent with that required for the formation of cellular structures in experiments \cite{Califano2008}. 

\section{Experimental Methods}

\textit{\textbf{Cell Culture}}: green florescent protein (GFP) expressing-human umbilical vein endothelial cells (HUVECs) (Angio-Proteomie) were expanded on 10mg/mL fibronectin-coated plates in Endothelial Cell Growth Medium-2 with BulletKit (EGM-2, Lonza). Cells used were between passages 3-12. Medium changes were performed every other day, and cells were split upon reaching ~80\% confluency.

\textit{\textbf{Polyacrylamide (PAA) fabrication}}: PAA hydrogels were fabricated similarly to previously published protocols \cite{Tse2010}. Briefly, hydrogels with relative stiffnesses (Young’s Modulus or elastic modulus, E) at ~1kPa, 2kPa, and 4.5kPa were fabricated by mixing acrylamide from 40\% stock solution (Sigma, A4058) with bis-acrylamide from 2\% stock solution (Sigma, M1533) in phosphate buffer saline (PBS). Air bubbles introduced during mixing were removed by vacuum gas-purge desiccation for 30min. The mixture was then mixed with 10\% ammonium persulfate (Sigma, A3426) and tetramethylethylenediamine (Sigma, T7024) at a 1:100 and 1:1000 ratios, respectively, initiating PAA polymerization. The PAA mixture was then sandwiched between an 18mm glass coverslip (Fisher) and a hydrophobically-treated, and dichlorodimethylsilane (Sigma, 440272)-coated glass slide. After 30min of PAA polymerization, the 18mm glass slide with the PAA hydrogel attached was carefully removed from the hydrophobic slide. Lastly, PAA hydrogels were functionalized with 0.2mg/mL sufosuccinimidyl-6-(4’-azido-2’-nitrophenylamino)-hexanoate (Pierce Biotechnology) followed by 10mg/mL fibronectin.

\textit{\textbf{Vascular Patterning}}: GFP-HUVECs were seeded on fibronectin- coated PAA hydrogels at a density of $2 \times 10^{4}$ cells/$cm^2$ and imaged after 16hrs on a Nikon Eclipse TE2000-U fluorescent microscope. The images were all processed using a custom-built image processing macro in FIJI2. 

\section{Image Analysis}
All image analysis used in this work was carried out using the open-source software ImageJ \cite{Schneider2012}.  Both experimental and simulation images (Fig.~\ref{fig:Figure3.5}a and Fig.~\ref{fig:Figure3.5}b, respectively) were imported into ImageJ and smoothed using “Gaussian Blur” at 4 pixels. Experimental background fluorescence removed with the “subtract background” function with a rolling ball radius of 300 pixels. The GFP-HUVECs were isolated using “Triangle” thresholding followed by an 8x dilation to preserve network connectivity which returned the image back into a binary image displaying the HUVEC morphology (Fig.~\ref{fig:Figure3.5}c) while the simulated networks are thresholded such that small scale features of assembly like compact rings are preserved while washing out the shape of the individual disks with no dilation factor(Fig.~\ref{fig:Figure3.5}d). At this point, both experimental and simulated images are binary. 

For calculating percolation and radius of gyration in experiments (Fig.~\ref{fig:Rg2}b), full simulation box is parsed into 73 non-overlapping subboxes and these subboxes are passed into a custom Python program which gives each pixel a label according to the cluster to which it belongs. If any two pixels have a separation distance greater than the width or height of the subbox and they belong to the same cluster - it is percolating configuration. The cluster whose label corresponds to the mode of the distribution is the largest cluster and radius of gyration is calculated for all the constituent pixels. 

For the fractal dimension in Fig.~\ref{fig:Figure4} , binary images are skeletonized with ImageJ's "Skeletonize" function and analyzed via the "Analyze Skeleton" tool to obtain branch length distributions. These branch length distributions allow us to calculate an average branch length for the experiments and simulations giving us a relative length scale between the two. This relative length scale allows us to sample the same relative space between simulation and experimental images. We then parse the simulation box into subboxes of size ($\text{simulation box length} \times \frac{\text{average branch length (experiment)}}{\text{average branch length (simulation)}}$) $\times$ ($\text{simulation box width} \times \frac{\text{average branch length (experiment)}}{\text{average branch length (simulation)}}$) (Fig.~\ref{fig:Figure3.5}c - insets). Each box is checked for local area fraction and then analyzed using ImageJ's "Fractal box count" tool with the default pixel array. For simulation network morphology metrics (Fig.~\ref{fig:Figure5}), circular markers are replaced by two markers along the dipole axis to pronounce anisotropy for ease of skeletonization.
 
\section*{Data Availability}
The raw data supporting the conclusions of this article will be made available by the authors, without undue reservation.


\section*{Acknowledgements}
PN was supported by a graduate fellowship funding from the National Science Foundation: NSF-CREST: Center for Cellular and Biomolecular Machines (CCBM) at the University of California, Merced: NSF-HRD-1547848. PN and KD would like to acknowledge support from the National Science Foundation (NSF-CMMI-2138672). AG would like to acknowledge support from the National Science Foundation (NSF-DMS-1616926). All authors would like to acknowledge support from the NSF-CREST: Center for Cellular and Biomolecular Machines at UC Merced (NSF-HRD-1547848) and the NSF Science and Technology Center for Engineering Mechanobiology awaard (NSF-CMMI-154857). 


\bibliography{bibliography}
 \bibliographystyle{unsrt}
 

\cleardoublepage
\appendix
\onecolumngrid
\section*{Supplementary Information (SI)}
\setcounter{figure}{0} 
\makeatletter
\renewcommand{\tablename}{Table S} 
\makeatletter

\renewcommand{\appendixname}{Supplementary Section} 
\renewcommand{\figurename}{Supp. Fig.}

\section{Elastic dipole interaction model}
\label{Model}

\begin{figure*}[h]  
		\centering
		\includegraphics[width=\textwidth]{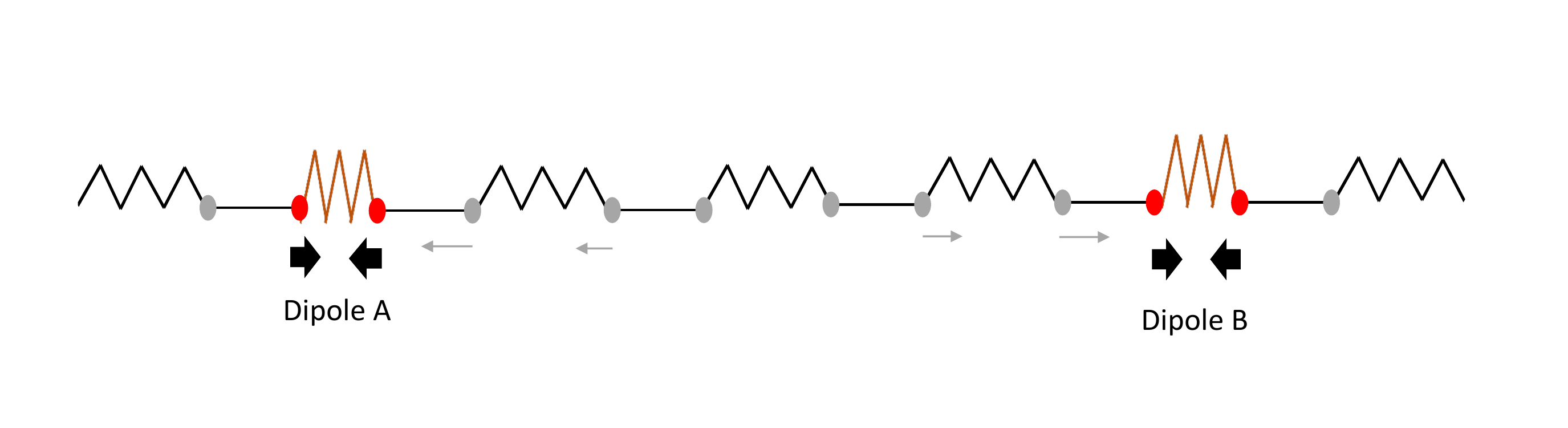}
		\caption{ 1D spring model illustrating origin of elastic interaction potential between two contractile dipoles.}
		\label{fig:1D} 
	\end{figure*}



\begin{figure*}[h]  
		\centering
		\includegraphics[width=\textwidth]{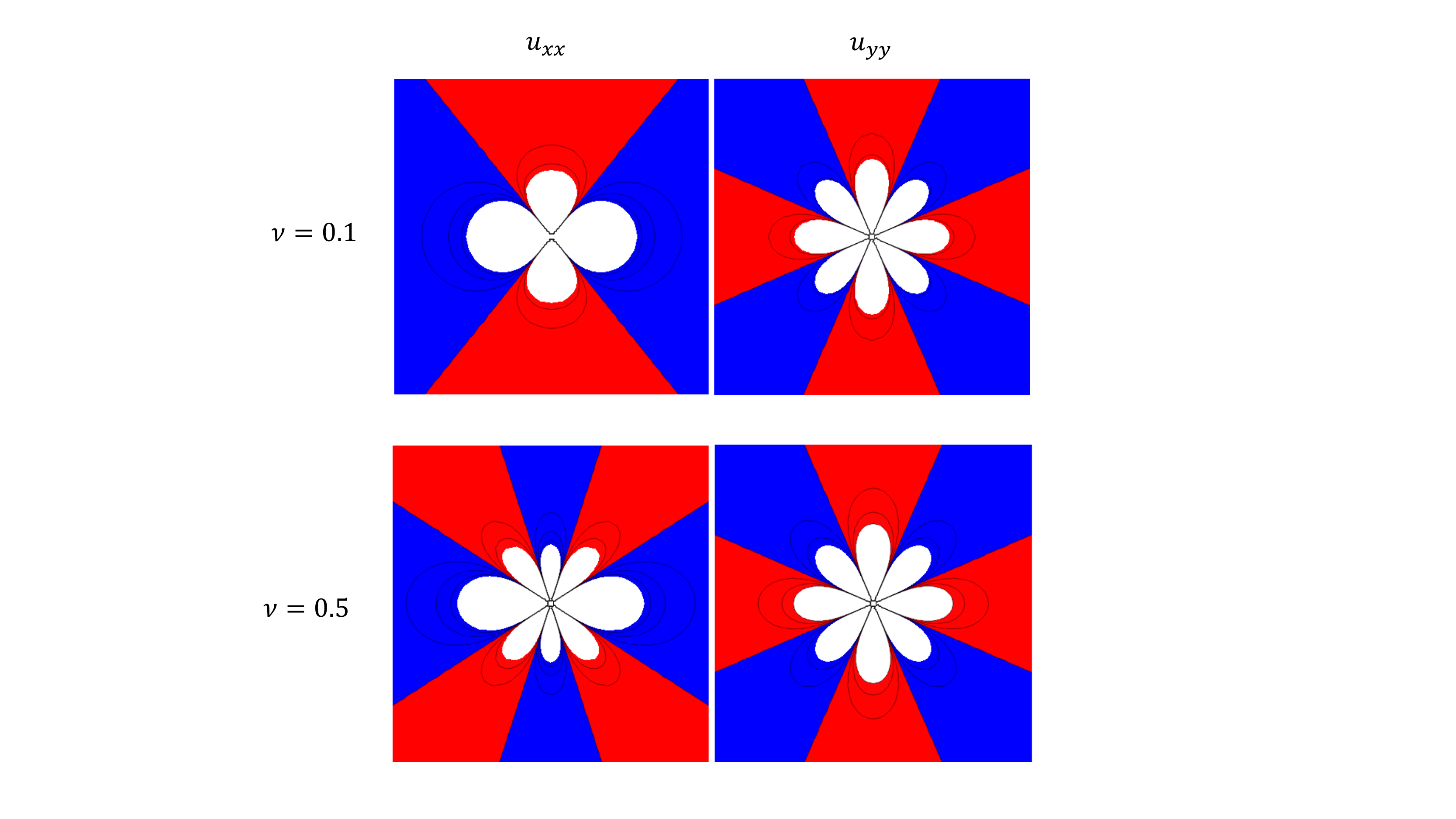}
		\caption{$u_{xx}$(left) and $u_{yy}$(right) components of strain field due to a contractile force dipole oriented along the $x-$axis in elastic half-space of a linear, isotropic medium. $u_{xx}$ component shows the $\nu=0.1$(top) map whose orientational distribution is that of an electric field from a quadrupole, while $\nu=0.5$(bottom) resembles an electric octupole. $u_{yy}$ has a similar structure for both shown values of Poisson's ratio, $\nu$.}
		\label{fig:Strains} 
	\end{figure*}

Consistent with adherent cell behavior on soft substrates, we assume our model cells are elongated and exert contractile traction forces at the poles of their long body axis. It is by this behavior that we model our cells as contractile force dipoles. The mechanical interaction between a pair of force dipoles is illustrated by the schematic in Supp. Fig. \ref{fig:1D} in the form of a 1D series of springs representing the effect of the elastic substrate. While the springs underlying the contractile dipoles are compressed, the springs between them are stretched. By moving to different positions in the medium for a given position of dipole $A$, the dipole $B$ can reduce the net substrate deformation energy by compressing regions stretched by dipole $A$. this physical interaction between elastic dipoles considered here is analogous to the interaction of an electric dipole with the electric field induced by another dipole. A similar reciprocal force results on dipole $A$, since the interactions are based on an elastic free energy. The physical origin of this force is the tendency of the passive elastic medium to minimize its deformations in response
to the active, contractile forces generated by the cells. We now assume these cells are on an isotropic, homogeneous, linear substrate in elastic halfspace of Young's modulus and Poisson's ratio $E$ and $\nu$, respectively, and derive the displacement field due to a coarse-graining of the traction forces on either side of the nucleus into single point like forces, $\mathbf{F}^{1}$ and $\mathbf{F}^{2}$ where $\mathbf{F}^{1} = -\mathbf{F}^{2}$, separated by a distance $a$. Let the center of the force distribution lie at $\mathbf{r}'$. Then, by elasticity theory, the displacement at position $\mathbf{r}$ can be written 
\begin{equation}
u_{i}(\mathbf{r})=G_{ij}(\mathbf{r}-(\mathbf{r}'-\frac{\mathbf{a}}{2}))F^{1}_{j}+G_{ij}(\mathbf{r}-(\mathbf{r}'+\frac{\mathbf{a}}{2}))F^{2}_{j},
\label{eq:displace}
\end{equation}
where $G_{ij}$ is the Green's function that captures the displacement in the elastic medium at the location of one cell (dipole) caused by the application of a point force at the location of the other \cite{landau_lifshitz_elasticity} defined as
\begin{equation}
G_{ij}({\bf r}) = \frac{1+\nu}{\pi E} \bigg[(1-\nu) \frac{\delta_{ij}}{r} + \nu \frac{r_i r_j}{r^3} \bigg].
\label{Boussinesq:Greensfunction}
\end{equation}
Replacing $\mathbf{F}^{1} = -\mathbf{F}^{2} = \mathbf{F} $ in eqn.~\ref{eq:displace} and performing a Taylor expansion about $\mathbf{r}-\mathbf{r}'$ to first order in $\mathbf{a}$ gives
\begin{equation}
u_{i}(\mathbf{r})=\partial_{k}G_{ij}(\mathbf{r}-\mathbf{r}')F_{j}a_{k} = \partial_{k}G_{ij}(\mathbf{r}-\mathbf{r}')P_{jk},
\end{equation}
where $P_{jk}=F_{j}a_{k}$ is the force dipole representation of one of our cell's force distribution, $\partial k$ is the partial derivative with respect to $x_{k}$, and terms of order $\mathbf{a}^2$ and higher have been neglected. We now write the strain by the derivatives of the displacement field as $u_{il}(\mathbf{r})=\frac{1}{2}(\partial_{l} u_{i}(\mathbf{r})+\partial_{i} u_{l}(\mathbf{r}))$, now substituting our symmetric Green's function, we can write the strain field as
\begin{equation}
u_{il}(\mathbf{r})=\partial_{l}\partial_{k}G_{ij}(\mathbf{r}-\mathbf{r}')P_{jk},
\end{equation}
where the $u_{xx}$ and $u_{yy}$ fields are shown in Supp. Fig.~\ref{fig:Strains}.
Lastly, we note that that by coupling the strain field due to one cell in the proximity of another, we can write the work done by deforming the medium and thus an effective pairwise interaction potential energy given by
\begin{equation}
W_{\alpha \beta}({\bf r}_{\alpha \beta}) =  P^{\beta}_{il} \partial_k \partial_l G^{\alpha \beta}_{ij}({\bf r}_{\alpha \beta}) P^{\alpha}_{jk},
\label{eq:elastic}
\end{equation}
where $P_{\alpha}$ and $P_{\beta}$ are the magnitude of the contractile force dipole exerted by cell $\alpha$ and cell $\beta$, respectively. $E$ is the Young's modulus of the elastic substrate,  
$\nu$ is Poisson's ratio, and ${\bf r}_{\alpha \beta} = \bm{r}_{\beta}-\bm{r}_{\alpha}$ is the separation vector connecting the positions of cell dipoles, $\alpha$ and $\beta$.

\begin{figure*}[h]  
		\centering
		\includegraphics[width=\textwidth]{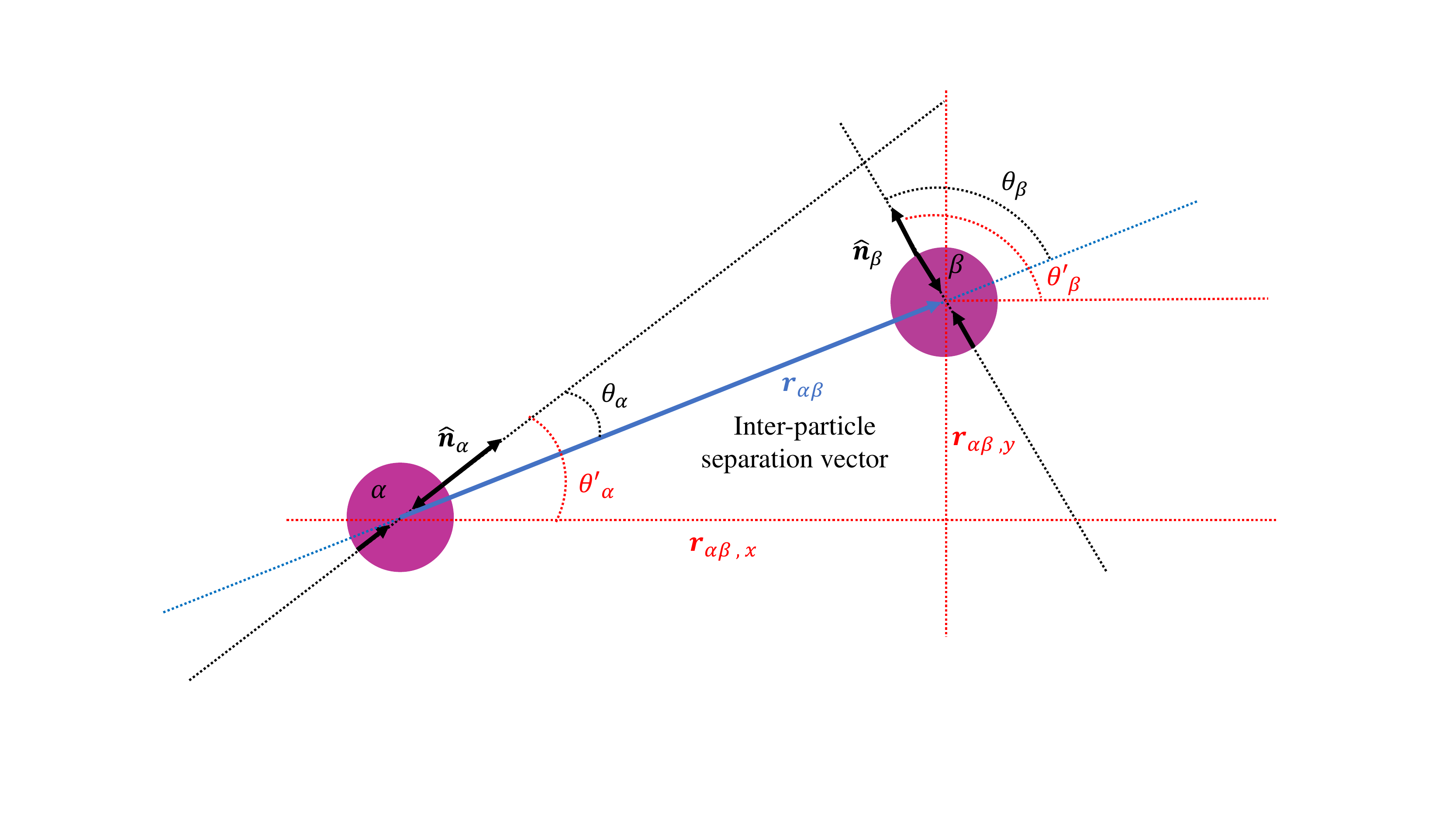}
		\caption{Schematic of two interacting particles with all relevant angles and vectors labeled. $\hat{\mathbf{n}}_{i}$ are unit vectors of force dipoles. $\theta'_{i}$ are angles of force dipoles with respect to the lab frame x-axis. $\theta_{\alpha}$ and $\theta_{\beta}$ are angles of force dipoles with respect to their separation vector $\mathbf{r}_{\alpha \beta}$ which has components $\mathbf{r}_{\alpha \beta , x}$ and $\mathbf{r}_{\alpha \beta , y}$.}
		\label{fig:Schematic2}
\end{figure*}

By transforming to the separation vector coordinate frame, the cell-cell elastic  potential can be written as \cite{Bischofs2004} 
\begin{equation}
W_{\alpha \beta} =  \frac{P_{\alpha}P_{\beta}}{ E r_{\alpha \beta}^{3}} f(\nu,\theta_{\alpha},\theta_{\beta}),
\label{ElInt}
\end{equation}
where $\cos(\theta_{\alpha}) = \bm{\hat{n}}_{\alpha}\cdot{\bm{r}_{\alpha \beta}}$ and $\cos(\theta_{\beta}) = \bm{\hat{n}}_{\beta}\cdot{\bm{r}_{\alpha \beta}}$ are the orientations of cell $\alpha$ and cell $\beta$ with respect to their separation vector, respectively. All relevant geometrical aspects of this interactions are realized and labeled in Supp. Fig. \ref{fig:Schematic2}. The dependence on these angles and the Poisson's ratio is collected in the function,
\begin{equation}
    f(\nu,\theta_{\alpha},\theta_{\beta}) = \frac{\nu(\nu+1)}{2\pi} \Big(3(\cos^{2}\theta_{\alpha}+\cos^{2}\theta_{\beta}-5\cos^{2}\theta_{\alpha}\cos^{2}\theta_{\beta}-\tfrac{1}{3})-(2-\nu^{-1})\cos^{2}(\theta_{\alpha}-\theta_{\beta})-3(\nu^{-1}-4)\cos\theta_{\alpha}\cos\theta_{\beta}\cos(\theta_{\alpha}-\theta_{\beta})\Big).
\end{equation}
For simplicity, We will assume the magnitude of all contractile cell force dipoles in our system are equal, so $P_{\alpha}=P_{\beta}=P$, which is justified when considering a culture of identical cells.



Taking derivatives of eqn. \ref{ElInt} with respect to $x_{\beta}$ and $y_{\beta}$ to compute the x and y components of the force, respectively, on cell $\alpha$ from cell $\beta$ yields

    \begin{align}
    - {\frac{dW_{\alpha \beta}}{dx_{\alpha}}} &= {\frac{dW_{\alpha \beta}}{dx_{\beta}}} = {\frac{\partial W_{\alpha \beta}}{\partial x_{\beta}}} + {\frac{\partial W_{\alpha \beta}}{\partial \theta_{\alpha}}}{\frac{\partial \theta_{\alpha}}{\partial x_{\beta}}} + {\frac{\partial W_{\alpha \beta}}{\partial \theta_{\beta}}}{\frac{\partial \theta_{\beta}}{\partial x_{\beta}}} \\ \nonumber
    &=-\frac{3P^{2}(1+\nu)}{16 \pi E r_{\alpha \beta}^{5}}\Big(-2+2\nu +6(\nu - 1)(\cos2\theta_{\alpha}+\cos2\theta_{\beta})+(\nu - 2)\cos2(\theta_{\alpha}-\theta_{\beta}) - 15\nu \cos2(\theta_{\alpha}+\theta_{\beta})\Big)\mathbf{r}_{\alpha \beta, x}
    \\ \nonumber
    &- \frac{P^{2}(1+\nu)}{16 \pi E r_{\alpha \beta}^{5}}\Big(12(\nu-1)(\sin2\theta_{\alpha}+\sin2\theta_{\beta})-60\nu\sin2(\theta_{\alpha}+\theta_{\beta})\Big)\mathbf{r}_{\alpha \beta, y}
    \end{align}
    
and

    \begin{align}
    -{\frac{dW_{\alpha \beta}}{dy_{\alpha}}} &= {\frac{dW_{\alpha \beta}}{dy_{\beta}}} = {\frac{\partial W_{\alpha \beta}}{\partial y_{\beta}}} + {\frac{\partial W_{\alpha \beta}}{\partial \theta_{\alpha}}}{\frac{\partial \theta_{\alpha}}{\partial y_{\beta}}} + {\frac{\partial W_{\alpha \beta}}{\partial \theta_{\beta}}}{\frac{\partial \theta_{\beta}}{\partial y_{\beta}}} \\ \nonumber
    &=-\frac{3P^{2}(1+\nu)}{16 \pi E r_{\alpha \beta}^{5}}\Big(-2+2\nu +6(\nu - 1)(\cos2\theta_{\alpha}+\cos2\theta_{\beta})+(\nu - 2)\cos2(\theta_{\alpha}-\theta_{\beta}) - 15\nu \cos2(\theta_{\alpha}+\theta_{\beta})\Big)\mathbf{r}_{\alpha \beta, y}\\ \nonumber
    &+ \frac{P^{2}(1+\nu)}{16 \pi E r_{\alpha \beta}^{5}}\Big(12(\nu-1)(\sin2\theta_{\alpha}+\sin2\theta_{\beta})-60\nu\sin2(\theta_{\alpha}+\theta_{\beta})\Big)\mathbf{r}_{\alpha \beta, x}\:,
    \end{align}
    
where $\mathbf{r}_{\alpha \beta, x} \equiv x_{\beta}-x_{\alpha}$ and $\mathbf{r}_{\alpha \beta, y} \equiv y_{\beta}-y_{\alpha}$ are the $x$ and $y$- components of the separation vector $\mathbf{r_{\alpha \beta}}$, respectively.

Similarly, for the torque on cell $\alpha$ by cell $\beta$, we take a derivative of the elastic potential with respect to $\theta_{\alpha}$ 
    \begin{equation}
    \begin{aligned}
    -{\frac{\partial W_{\alpha \beta}}{\partial \theta_{\alpha}}} = -\frac{P^{2}(1+\nu)}{8 \pi E r_{\alpha \beta}^{3}}\Big(-6(\nu - 1)\sin2\theta_{\alpha} - (\nu - 2)\sin2(\theta_{\alpha}-\theta_{\beta}) + 15\nu\sin2(\theta_{\alpha}+\theta_{\beta})\Big)\:.
    \end{aligned}
    \end{equation}

\section{Nondimensionalization of Langevin equations}
\setcounter{equation}{0}
We begin with the Langevin equation for cell position stated on the first line of the Model section of the main text.

\begin{equation}
    {\frac{d\mathbf{r}_{\alpha}}{dt}} =  -\mu_{T}\sum_{\beta} \frac{\partial W_{\alpha \beta}}{\partial {\bf r}_{\alpha}} + \sqrt{2D_{\mathrm{T}}}\:\: \bm{\eta}_{\alpha,\mathrm{T}}(t)\:,
    \end{equation}
where $D_{\mathrm{T}}$ is the effective translational diffusivity quantifying the random motion of an isolated moving cell, with $\bm{\eta}$ as a random white noise term whose components satisfy 
$\langle \eta_{i}(t) \eta_{j}(t') \rangle = \delta(t-t') \delta_{ij}$. Note that $\eta$ - the noise term describing active cell motility - has units of $t^{-1/2}$. $W_{\alpha \beta}$ is a long-range elastic potential (full form shown in kd{write equation number}) when $\sigma \leq \mathrm{r_{\alpha \beta}} \leq 3\sigma$ and a steric spring given by $W_{\alpha \beta} = {\frac{1}{2}}k {(\sigma - r_{\alpha \beta})^2}$ when $0 <\mathrm{r_{\alpha \beta}} \leq \sigma$.
    
We now choose characteristic time, length, and energy scales. Let $\mathbf{r^{*}} = \frac{\mathbf{r}}{\sigma}$ be  a dimensionless distance vector scaled by cell size, let  $W^{*}_{\alpha \beta} = \Big(\frac{P^{2}}{16E \sigma^{3}}\Big)^{-1}W_{\alpha \beta}$ be a dimensionless energy scaled by elastic energy at cell length separation, and let $t^{*} = \frac{P^{2}\mu_{\mathrm{T}}}{16E\sigma^{5}}t$ be a dimensionless time scaled by an elastic interaction.

Non-dimensionalizing our translational Langevin equation using the above characteristic scales gives us the following equation
    \begin{equation}
    {\frac{d\mathbf{r^{*}_{\alpha}}}{dt^{*}}} =  -\sum_{\beta} \frac{\partial W^{*}_{\alpha \beta}}{\partial {\mathbf{r^{*}_{\alpha}}}} + \sqrt{\frac{2}{A}}\:\: \bm{\eta}^{*}_{\alpha,\mathrm{T}}(t^{*})\:,
    \end{equation}
    
where 
    \begin{equation}
    A \equiv \frac{P^{2}\mu_{\mathrm{T}}}{16E\sigma^{3}D_{\mathrm{T}}}=\frac{P^{2}}{16E\sigma^{3}k_\mathrm{B}T_{\mathrm{eff}}} = \frac{\mathcal{E}_{c}}{k_\mathrm{B}T_\mathrm{eff}}
    \end{equation}
    is a dimensionless parameter that is the ratio of characteristic elastic energy to an effective temperature called the effective elastic interaction.
    
The Langevin equation for cell orientation is given by
    \begin{equation}
    {\frac{d\hat{\mathbf{n}}_{\alpha}}{dt}} =  -\mu_{R} \sum_{\beta}\Big(\hat{\mathbf{n}}_{\alpha}\times\frac{\partial W_{\alpha \beta}}{\partial \hat{\mathbf{n}}_{\alpha}}\Big) + \sqrt{2D_{\mathrm{R}}}\:\: {\bm{\eta}}_{\alpha,\mathrm{R}}(t)
    \label{orientational_LE}
    \end{equation}
   where $\hat{\bm{n}}$ is the cell orientation and $D_{\mathrm{R}}$ is the effective rotational diffusivity quantifying the random reorientations of an isolated moving cell.

Nondimensionalizing eqn.~\ref{orientational_LE} with the same scales as in the translational Langevin equation and assuming  $\mu_{\mathrm{R}}\sigma^{2}=\mu_{\mathrm{T}}$ and $D_{\mathrm{R}}\sigma^{2}=D_{\mathrm{T}}$ gives us
    
    \begin{equation}
    {\frac{d\hat{\mathbf{n}}_{\alpha}}{dt^{*}}} =  - \sum_{\beta}\Big(\hat{\mathbf{n}}_{\alpha}\times\frac{\partial W^{*}_{\alpha \beta}}{\partial \hat{\mathbf{n}}_{\alpha}}\Big) + \sqrt{\frac{2}{A}}\:\: {\bm{\eta}}^{*}_{\alpha,\mathrm{R}}(t^{*})\:.
    \end{equation}
    
\section{Numerical methods} \label{App C}
\setcounter{equation}{0}

We can rewrite the nondimensionalized Langevin equations obtained in Appendix B in a discrete form as follows:
    \begin{equation}
    \mathbf{r^{*}_{\alpha}}(t+\Delta t) = \mathbf{r^{*}_{\alpha}}(t) -\sum_{\beta} \frac{\partial W^{*}_{\alpha \beta}}{\partial {\mathbf{r^{*}_{\alpha}}}}\Delta t + \sqrt{\frac{2}{A}}\:\: \bm{\eta}^{*}_{\alpha,\mathrm{T}}\sqrt{\Delta t}\:,    
    \end{equation}
and    
    \begin{equation}
    \theta'_{\alpha}(t+\Delta t) = \theta'_{\alpha}(t) - \sum_{\beta}\frac{\partial W^{*}_{\alpha \beta}}{\partial \theta'_{\alpha}}\Delta t + \sqrt{\frac{2}{A}}\:\: {\eta}^{*}_{\alpha,\mathrm{R}}\sqrt{\Delta t}\:,    
    \end{equation}

where $\theta'_{\alpha} \equiv \tan^{-1} \frac{y_{\alpha}}{x_{\alpha}} + \theta_{\alpha}$ is the angle between $\mathbf{\hat{n}}_{\alpha}$ and the lab frame x-axis. A schematic of all the variables used is shown in Supp. Fig.~\ref{fig:Schematic2}.  The time step used in the simulations is $\Delta t = .000625$. Each cell is initialized at a random orientation and position inside the simulation box - a square with length $L$. The position and angle of each cell is updated at every interval $\Delta t$ according to equations C1 and C2 with periodic boundary conditions. The simulations are run for $1 \times 10^{5}$ time steps with annealing of the effective temperature to avoid metastable states and to promote cell activity before many contacts form. The cells are kept at the final effective temperature for $1/4$ the total simulation time which equates to roughly twenty four hours of experimental time, an appropriate time after which to report cell configurations.



\section{Phase portrait of simulation final snapshots on compressible and incompressible substrates}
\begin{figure*}[!h]  
		\centering
		\includegraphics[width=\textwidth]{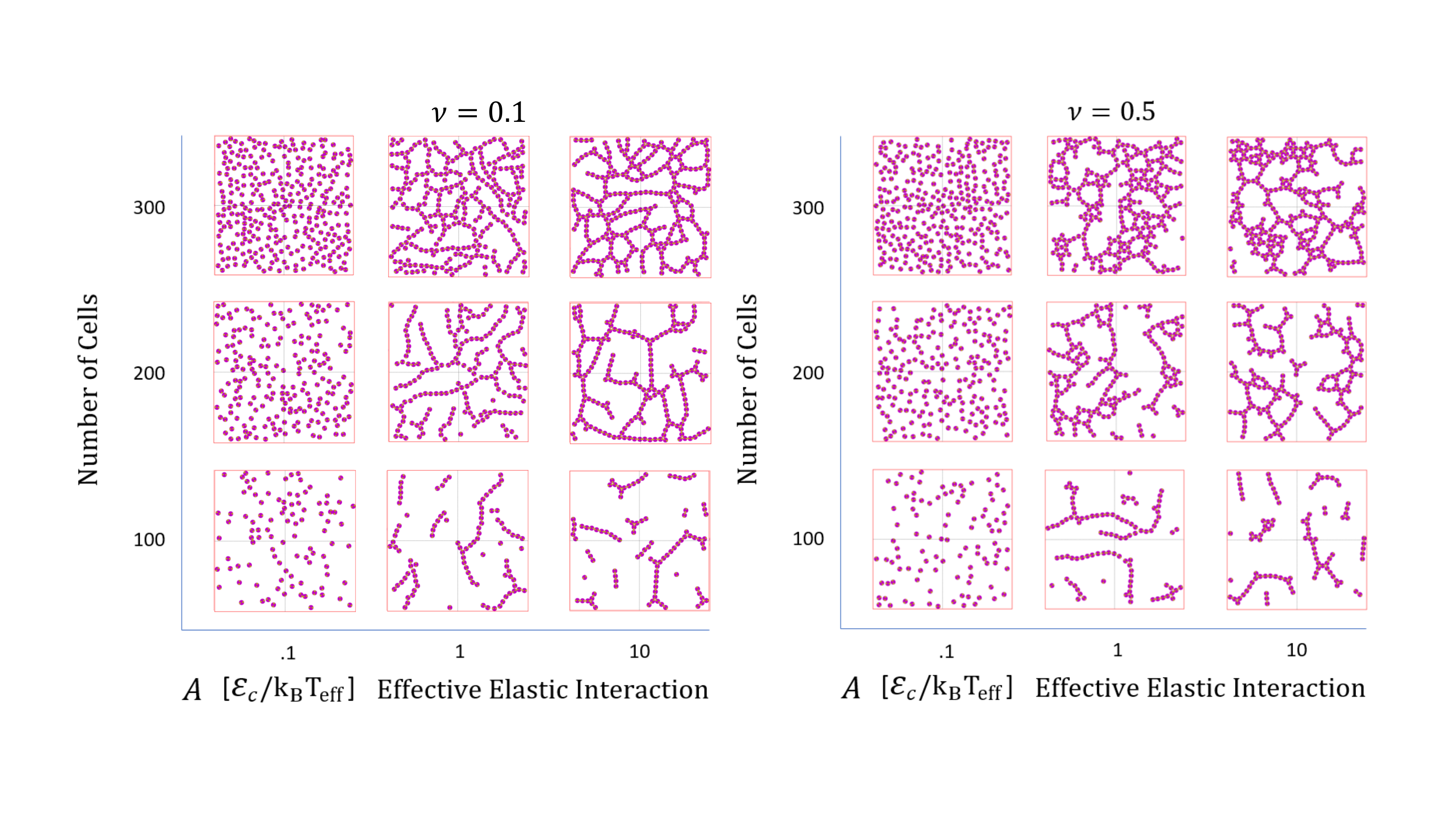}
		\caption{Simulation snapshots of final configurations in the parameter space of number of cells and $A \equiv \tfrac{\mathcal{E}_{c}}{k_\mathrm{B}T_{\mathrm{eff}}}\:$, the ratio of the characteristic elastic interaction strength and noise, for $\nu = 0.1$ (left) and $\nu = 0.5$ (right). At lower packing fractions, cells form segments of branches and stems. At lower A values, cells remain isolated. At higher values of A with sufficient packing fraction, cells form space spanning network configurations characterised by rings, branches, and junctions. At higher packing fractions, parallel chains occur frequently in these networks.}
		\label{fig:Portrait} 
	\end{figure*}

Fig.~\ref{fig:Portrait}b is shown in the main text. We now show the phase portrait for the low $\nu$ case in Supp. Fig.~\ref{fig:Portrait}a. While the trends and general dependence on $A$ and $N$ is the same for both values of Poisson's ratio, we can see from the $A=10, N=300$ cases that $\nu=0.1$ forms longer chained, larger ringed structures than the $\nu=0.5$ system which forms compact structures of tight rings and many junctions.

\section{Computational Analysis of Networks}
\setcounter{equation}{0}
\label{CompNet}
\subsection*{Identifying clusters} Each cell is assigned to a cluster by assigning an initial cell to zeroth cluster. 
 Then the cells in its neighbor list - a list identifying all other cells that are within $1.2\sigma$ of the central cell - are assigned to this cluster. The neighbor list of each of these neighbors are assigned this cluster label in an identical way. Once all neighbor lists have been exhausted, we search for unassigned cells and repeat the process with an incremented cluster number until every cell belongs to one or the other cluster. 
 Once each cell belongs to a cluster, cell-cell distances are checked. If the distance between any two cells within the same cluster is greater than or equal to the size of the simulation box, we consider that realization of the network to be percolating. This calculation is done at the final time step of forty simulations per data point shown in Fig. 3 for dipoles and ten simulations per data point for diffusive sticky disks. The average value and corresponding error are then plotted as a function of packing fraction $\phi$ in Fig.~3a and of the effective elastic interaction parameter $A$ in Fig.~3b.
\subsection*{Identifying junctions/branches} Final configurations of cells, like those shown in Figs. 1 and 2, are re-plotted with elongated black markers on cell positions along the direction of the dipole axis. This gives us networks like those shown in Fig.~6. These images are imported into imageJ \cite{Schneider2012}, Gaussian blurred, intensity thresholded, binarized, and skeletonized. By then using the "Analyze Skeleton" plugin in ImageJ, we obtain skeleton information including the full branch length distribution and junction counts \cite{ArgandaCarreras2010}.
\subsection*{Identifying rings} Instead of using the "Analyze Skeleton" plugin in ImageJ, we invert the binarized image described above and utilize the "Analyze Particles" function of ImageJ to obtain a distribution of rings and ring areas in the networks.


\section{Critical packing fraction dependent on box size}
\begin{figure*}[!h]  
		\centering
		\includegraphics[width=.8\textwidth]{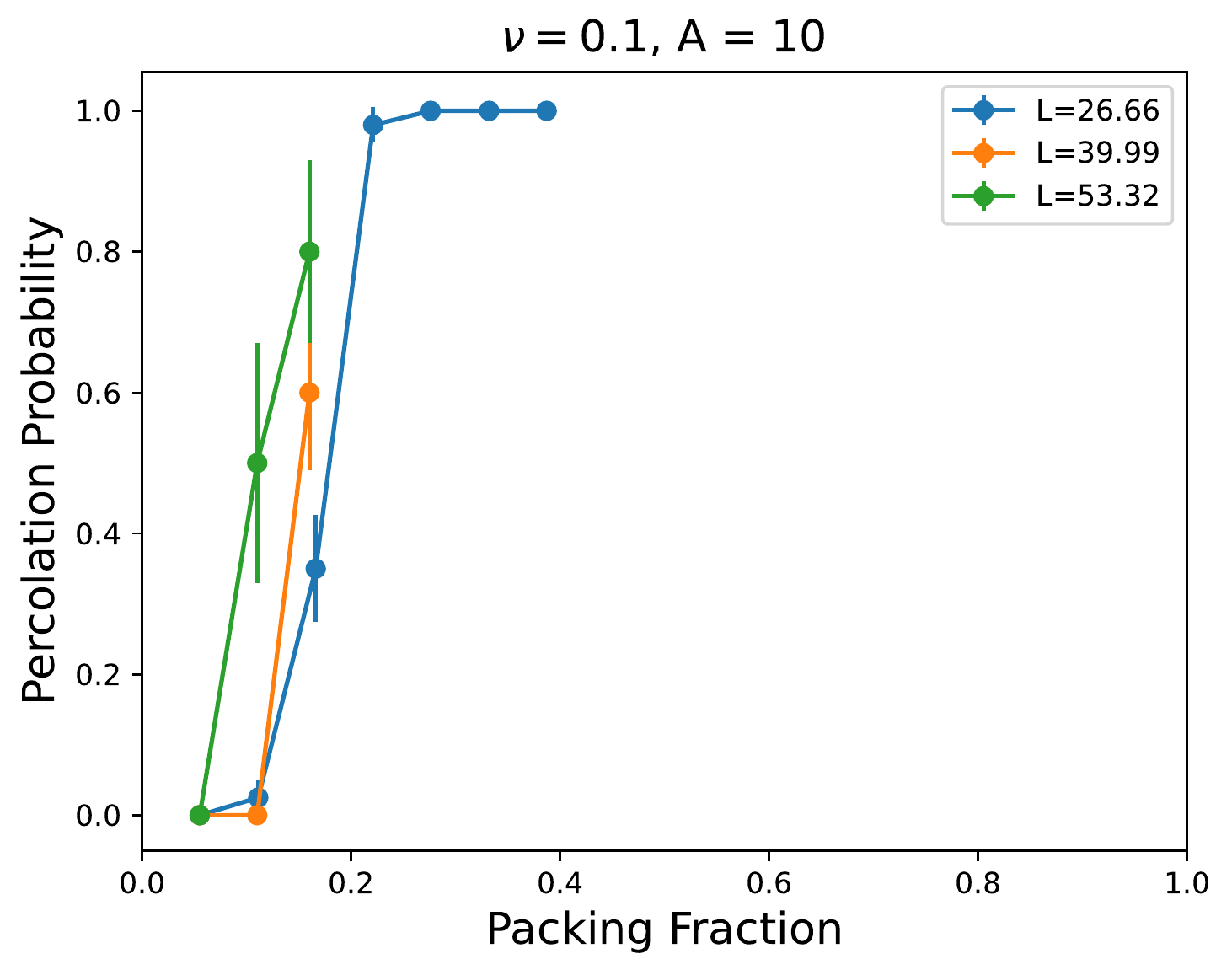}
		\caption{Critical packing fraction of elastic dipoles decreases with increasing box size. Due to the highly anisotropic nature of elastic dipolar interactions, dipoles will percolate at lower critical area fraction as (near the transition) area scales as $L^{2}$ whereas cluster size scales as $L^{d_{f}}$ where $d_{f}$ is the fractal dimension. Thus, the critical packing fraction will go as $L^{d_{f}-2}$ where $d_{f}-2 < 0$. }
		\label{fig:BoxSizePerc} 
	\end{figure*}

\setcounter{equation}{0}
In the main text, the connectivity percolation we report is for a specific box size $L=26.66\sigma$. This curve is subject to shift and/or dilate/contract under varying the box size (Supp. Fig.~\ref{fig:BoxSizePerc}). The box size we chose to use in the main text is appropriate (given our characteristic length scale of $\approx 50 \mu m$ to compare to in vitro experiments of cells on compliant substrates. 
\label{BoxSizePerc}

\section{Mapping the effective interaction parameter to substrate stiffness}
\setcounter{equation}{0}
\label{AtoE}

\begin{figure*}[h]  
		\centering
		\includegraphics[width= 0.7\textwidth]{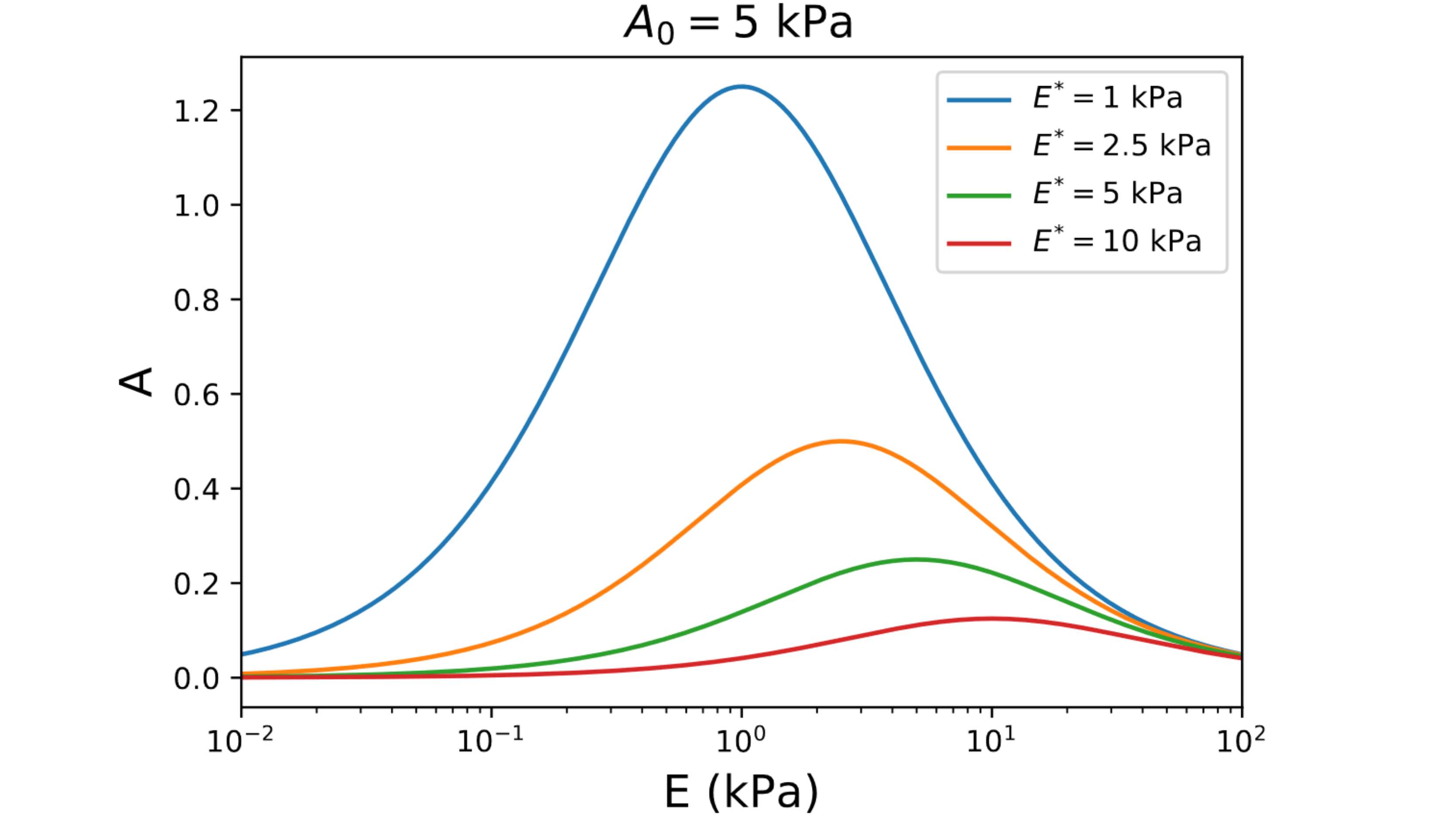}
		\caption{Mapping $A$ to $E$. Four curves characterized by various choices of optimal substrate stiffness $E^{*}$ for $A_{0}$ are shown. Range of $A$ values mapped to $E$ decreases as the choice of optimal stiffness increases.}
		\label{fig:AtoE} 
	\end{figure*}
	
\begin{figure*}[h]  
		\centering
		\includegraphics[width= 0.7\textwidth]{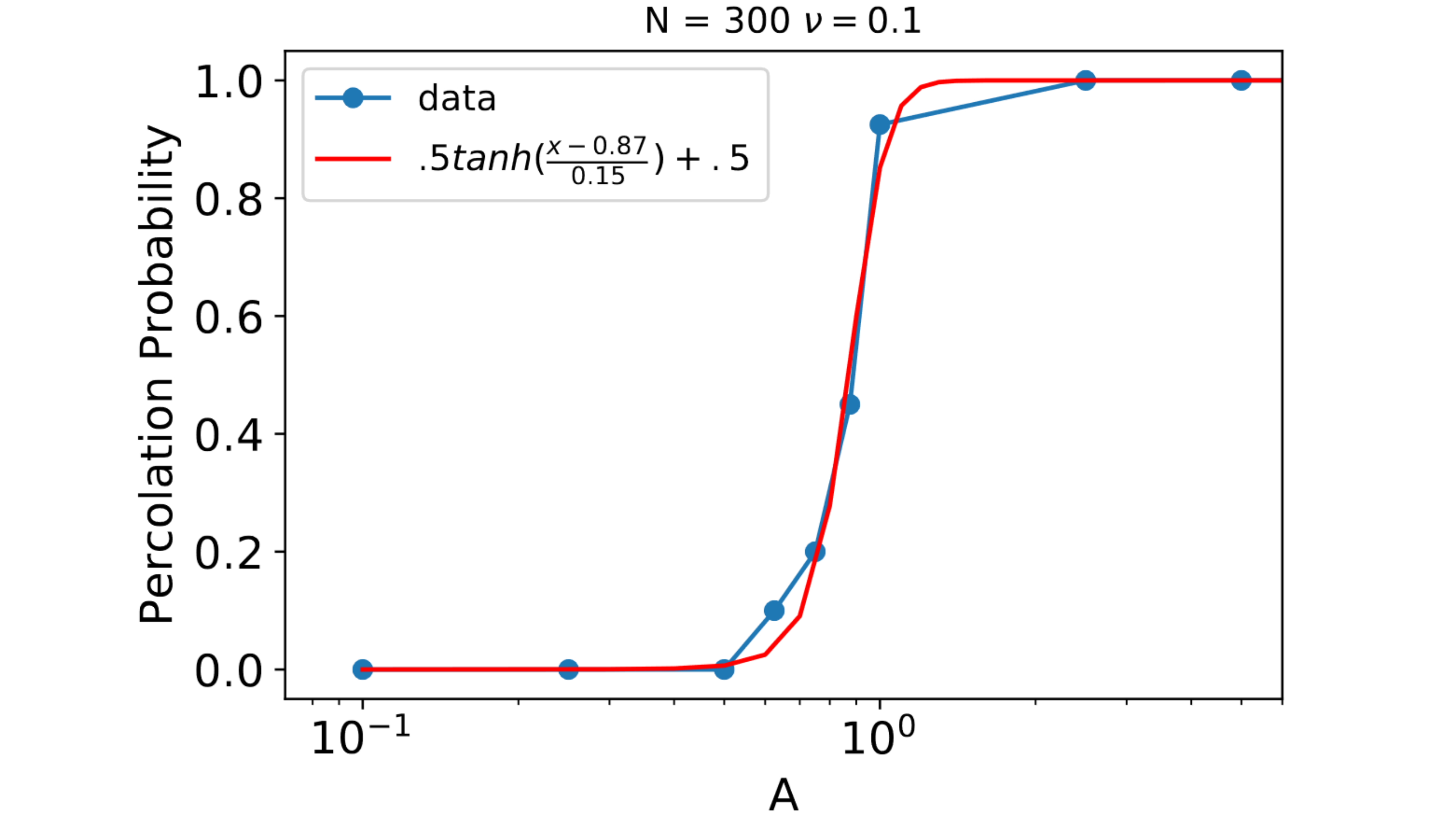}
		\caption{Percolation probability as a function of effective elastic interaction for $N=300 (\phi \approx .33)$ cells where $\nu=0.1$. The percolation curve is fit well to a hyperbolic tangent function with two free parameters corresponding to the position and width of the transition.}
		\label{fig:tanh_fit} 
	\end{figure*}	
	
	\begin{figure*}[h]  
		\centering
		\includegraphics[width=\textwidth]{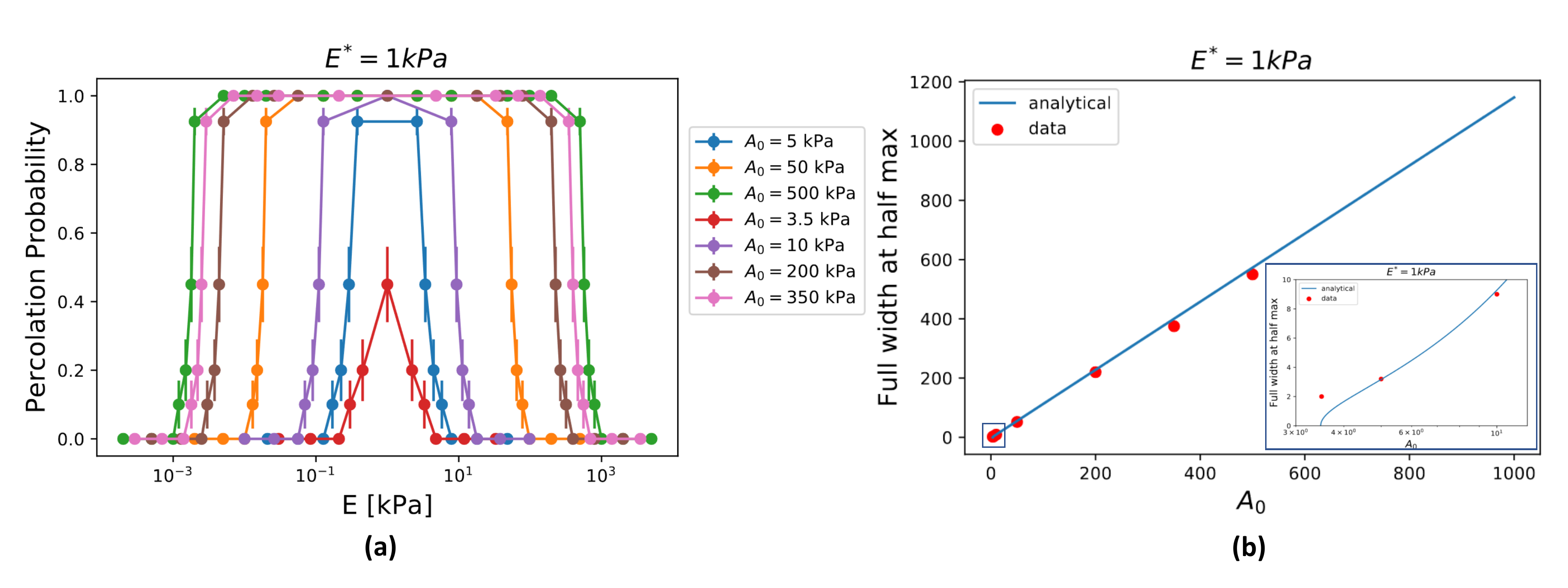}
		\caption{Percolation peak width as a function of $A_{0}$. (b) Percolation probability as a function of substrate stiffness for various values of $A_{0}$. (b) Blue curve shows the analytic expression for percolation peak width quadratic in $A_{0}$ gives good agreement with  mappings from simulation data, shown in red dots obtained from (a), except at lower values of $A_{0}$ where the analytical expression breaks down due to an assumption of transition $(p_{max} \approx 1$).}
		\label{fig:FWHM} 
	\end{figure*}

We have considered the effective elastic interaction, $A$, to be the model parameter which encodes stochasticity, cell forces, and substrate stiffness. We wish now to relate this parameter to an easily accessible and measurable experimental value - substrate stiffness. By  casting $A$ in terms of substrate stiffness, we aim to predict trends with varying substrate stiffness, which can be directly tested in experiment. It is known from  traction force experiments (such as Ref.~\cite{Ghibaudo2008}) that cells on elastic substrates adapt their forces to substrate stiffness. Adherent cells on softer substrates build fewer and smaller focal adhesions.  With increasing substrate stiffness, cells spread more and exert stronger traction forces which saturate to a constant value beyond a typical substrate stiffness $E^{*}$ which depends on cell type (table \ref{tab2}).
\begin{table}[!ht]
\caption{Contractility and optimal stiffness of various cell types.}
\label{tab2}
\begin{center}
\begin{tabular}{lll}
\toprule
\textbf{\boldmath{$P_{0}$} (J)}	& \textbf{\boldmath{$E^{*}$} (kPa)}	 & \textbf{Cell Type}\\
$10^{-12}$\cite{Bischofs2005}\cite{Bischofs2006} & $1{\text -}10$\cite{Califano2008, Rudiger2020} & Endothelial    \\
$10^{-13}$\cite{Hyland2014} & $0.1$\cite{SOLON20074453, Flanagan2002}  & Neuron    \\
$10^{-9}$\cite{mcb.2019.06415} & $20$\cite{SOLON20074453} & Smooth Muscle \\
$10^{-11}$ \cite{Bizanti2021} &         $10$ \cite{Bizanti2021} & Astrocyte \\
\end{tabular}
\end{center}
\end{table}
This mechanical adaptivity of the cell traction is modeled by considering a force dipole magnitude that scales with substrate stiffness as \cite{zemel_10},
    \begin{equation}
    P(E) = P_{0} E/(E + E^{\ast}). 
    \label{Eq:DipoleMagnitude}
    \end{equation}
Plugging in \ref{Eq:DipoleMagnitude} to our definition of $A$ gives us $A = \frac{P^{2}}{16E\sigma^{3}k_\mathrm{B}T_{\mathrm{eff}}} = \frac{P_{0}^{2}E}{16(E+E^{*})^{2}\sigma^{3}k_\mathrm{B}T_{\mathrm{eff}}}$, which has a non-monotonic dependence on substrate stiffness, as seen in Supp. Fig.~\ref{fig:AtoE}, reaching a maximum of $\frac{A_{0}}{4}$ where $A_{0} \equiv \frac{P_{0}^{2}}{16\sigma^{3}k_\mathrm{B}T_{\mathrm{eff}}}$ at $E=E^{\ast}$. We now have a mapping from effective interaction parameter $A$ to substrate stiffness $E$ which we can directly relate to experiments.


We know from experiments of endothelial cells on elastic substrates, that cells can form networks or remain isolated from one another depending on the substrate rigidity \cite{Califano2008}. This is to say that there is a range of viable substrate stiffnesses over which cells will self-assemble into vascular networks. We now ask if we can predict the viable range of stiffnesses that accommodate network formation. We use the metric of percolation probability to quantify the tendency for network formation. We know that we can make these predictions numerically as we now have a mapping from $A$ to $E$ for our simulations. Percolation vs. substrate stiffness curves are shown in Supp. Fig.~\ref{fig:FWHM}b for various values of $A_{0} \equiv \frac{P_{0}^{2}}{16k_\mathrm{B}T_\mathrm{eff}\sigma^{3}}$, the effective elastic interaction without stiffness dependence.

We now seek to develop an analytic treatment of the substrate dependent percolation metric and obtain a closed form expression which predicts the range of substrate stiffnesses conducive to network formation. Supp. Fig. \ref{fig:tanh_fit} shows that percolation vs. A is well fit by a hyperbolic tangent function with two fit parameters $A^{*}$ and $k$ such that 
\begin{equation}
    p = .5\tanh\Big(\frac{A-A^{*}}{k}\Big)+.5,
    \label{Eq:analyticexpression}
    \end{equation}
where $p$ is the percolation probability. We will use the full width at half maximum (FWHM) to represent the range of values over which networks are formed. In order to find the FWHM, which we will call $\Sigma$, of the above function, we set $p$ equal to half of its maximum value, which we assume is $1$. This gives us the condition $A = A^{\ast}$,
which we can rewrite in the following way
\begin{equation}
    E^{2}+\Big(2E^{*}-\frac{P_{0}^{2}}{16\sigma^{3}k_\mathrm{B}T_\mathrm{eff}A^{\ast}}\Big)E+E^{*2}=0.
    \end{equation}
Thus, we obtain an expression for the FWHM of our percolation curves given by
\begin{equation}
    \Sigma = \sqrt{\Big(\frac{P_{0}^{2}}{16\sigma^{3}k_\mathrm{B}T_{\mathrm{eff}}A^{\ast}} - 2E^{\ast}\Big)^{2}-4E^{\ast2}}.
    \label{Eq:FWHM2}
    \end{equation}
Supp. Fig. \ref{fig:FWHM}a shows a comparison of peak width for various values of $A_{0}$ computed with \ref{Eq:FWHM2} and computed numerically from Supp. Fig.~\ref{fig:FWHM}b. The plot shows great agreement between the analytical prediction and numerical results except at values of $A_{0}$ that are close to the analytical solution condition $\frac{P_{0}^{2}}{16\sigma^{3}k_\mathrm{B}T_{\mathrm{eff}}A^{*}} \geq 4E^{*}$. This disparity, shown in the inset of Supp. Fig.~\ref{fig:FWHM}a is due to the assumption that the maximum percolation value is $1$, which is not the case for the red curve in Supp. Fig.~\ref{fig:FWHM}b.

Thus, \ref{Eq:FWHM2} provides us a closed form expression, valid over a large interval of parameter space, for the range of substrate stiffnesses over which cells will form percolating networks. This value is dependent on cell forces, effective temperature, cell size, and a fit parameter $A^{*}$ which represents the position of the percolation transition.  In particular, it predicts that higher force dipole magnitude ($P_{0}$), lower noise ($T_{\mathrm{eff}}$), and higher cell density corresponding to lower required elastic interaction for percolation ($A^{\ast}$), all lead to wider peaks in percolation vs substrate stiffness.

\section{Junction count shows similar behavior to neighbor counts}
\setcounter{equation}{0}
\begin{figure*}[!ht]  
		\centering
		\includegraphics[width=\textwidth]{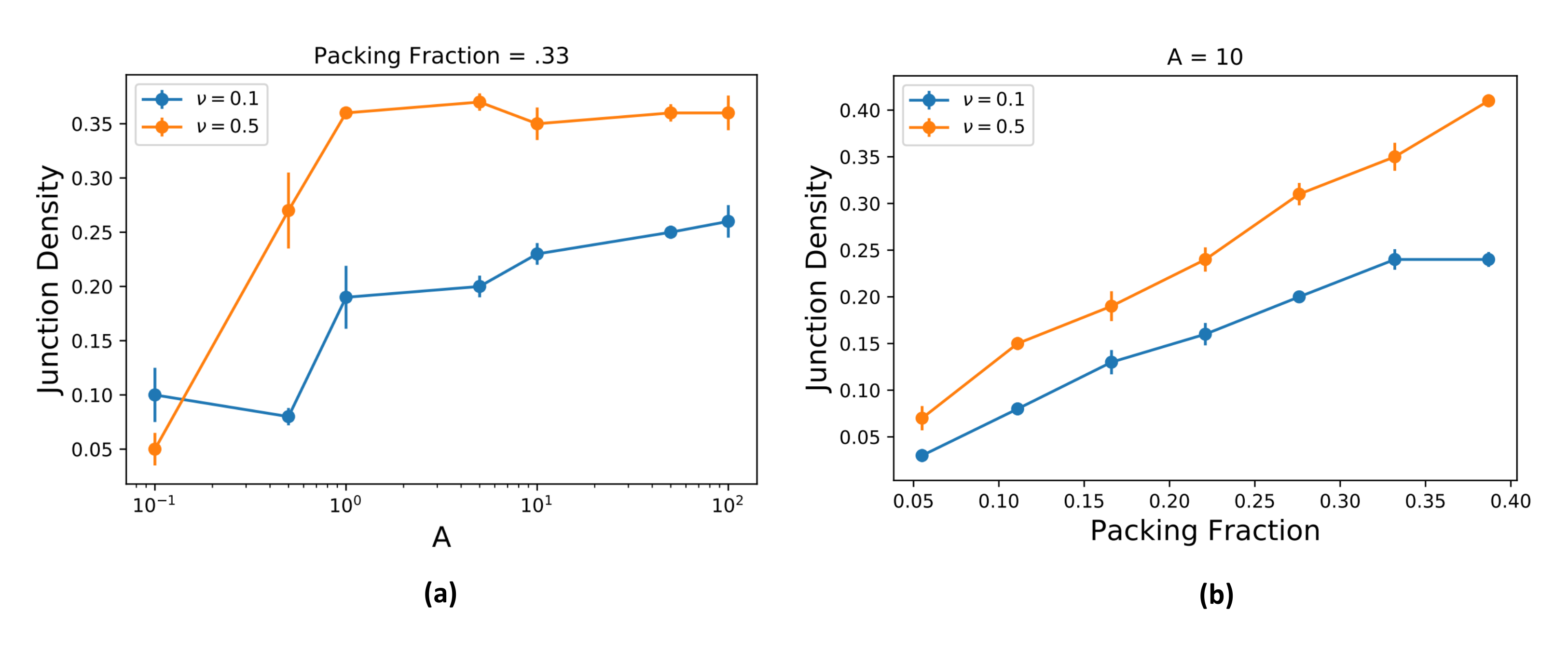}
		\caption{Junction density shows a trend similar to neighbor counts. (a) Junction density vs. $A$ shows at low $A$, few cells are part of a junction. As $A$ increases, irreversible networks structures are formed where the $\nu=0.5$ systems exhibit a greater capacity to form junctions - structures which produce greater neighbor counts. (b) Junction density vs. $\phi$ (or $N$) increases as a function for all $N$ when $\nu=0.5$. Junction density begins to decrease at highest $\phi$ as the system begins to form more parallel strings in the low $\nu$ case.}
		\label{fig:JuncDensity} 
	\end{figure*}
Another metric that can be used to probe the morphologies of our branched networks is junction density - the number of cells connected to a node after skeletonizing normalized by the total number of cells. Junction density is shown in Supp. Fig.~\ref{fig:JuncDensity} as a function of $A$ (left) and $\phi$ (or $N$) (right).  Unsurprisingly, junction density vs. $A$ follows the same trend as neighbors vs. $A$ shown in Fig. 6e in the main text as junctions are structures which promote higher neighbor counts. In contrast with neighbors vs. $\phi$, our highest value of packing for junction density exhibits a different trend. While neighbor counts continue to increase, junction density slightly decreases for $\nu=0.1$. This is due to the ground states explored in the previous section. At of our highest packing fraction, we begin to transition out of the dilute regime where we know $\nu=0.1$ dipoles will form parallel strings. Tightly packed parallel strings will have a high neighbor count but no junctions. 

\section{Dependence of percolation on packing fraction and elastic interactions}
\label{FullPerc}
\begin{figure*}[!h]  
		\centering
		\includegraphics[width=\textwidth]{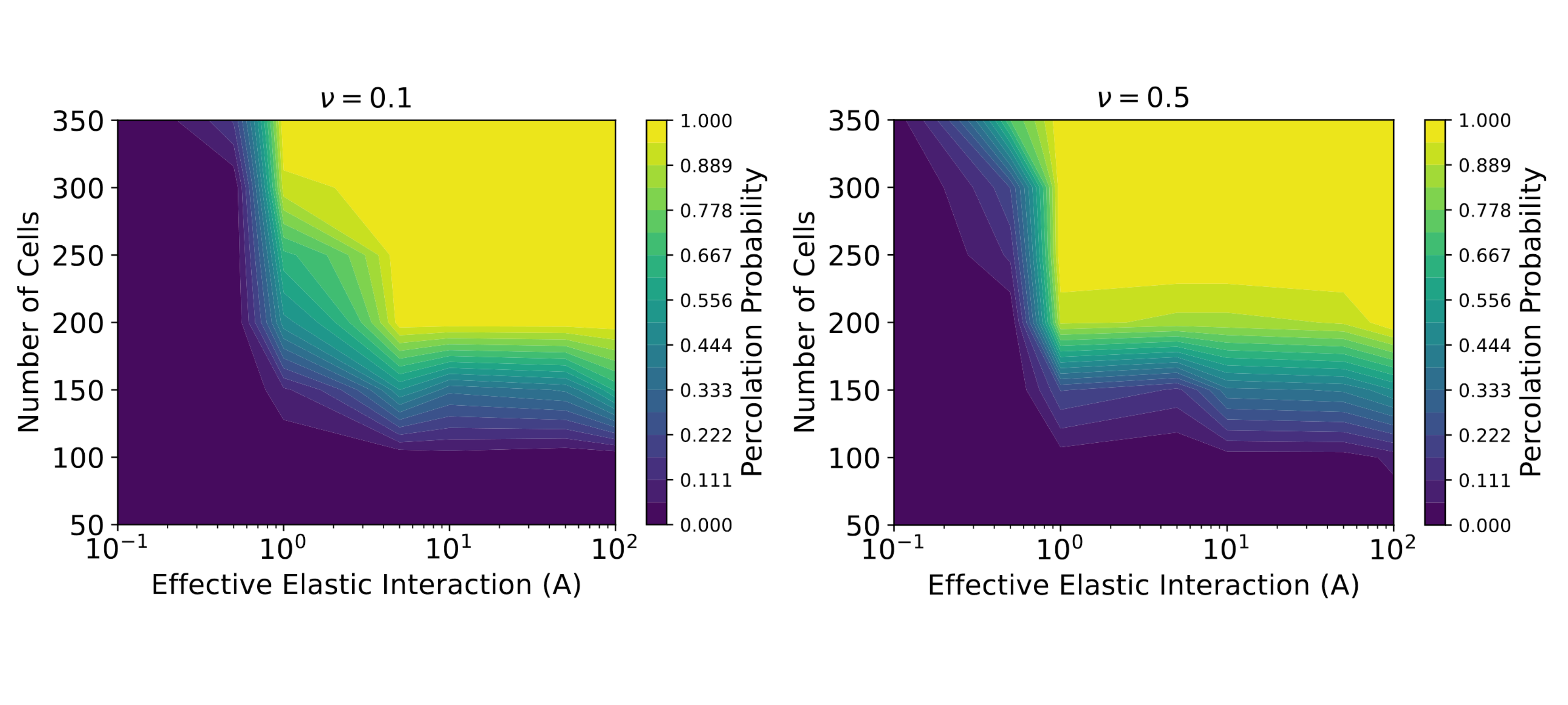}
		\caption{Percolation contour plots show $\nu=0.1$ is more efficient with respect to $N$ while $\nu=0.5$ is more efficient with respect to $A$. (a) Color represents percolation probability in ($A$,$N$) space for $\nu=0.1$ (left) and $\nu=0.5$ (right). For $A<5$, $\nu=0.5$ is more percolating while for $A \geq 5$, $\nu=0.1$ is more percolating.}
		\label{fig:PercMap} 
	\end{figure*}
While we report several percolation curves in Fig.3 of the main text, Supp. Fig.~\ref{fig:PercMap} shows a percolation contour map in ($A$,$N$) space. At low $A$, $\nu=0.5$ percolates more reliably than the low $\nu$ counterpart as the system is more resilient to noise. At $A \geq 5$, however, $\nu=0.1$ percolates more reliably with fewer cells as we have seen the lower $\nu$ system forms more extended structures in general. The full percolation map then shows that networks at low Poisson's ratio are more efficient with respect to number of cells, whereas the high Poisson's ratio networks more efficient with respect to effective elastic interaction.

\section{Irrigation area reveals a marginal efficiency for networks at low Poisson's ratio}
\setcounter{equation}{0}
\begin{figure*}[!h]  
		\centering
		\includegraphics[width=0.9\textwidth]{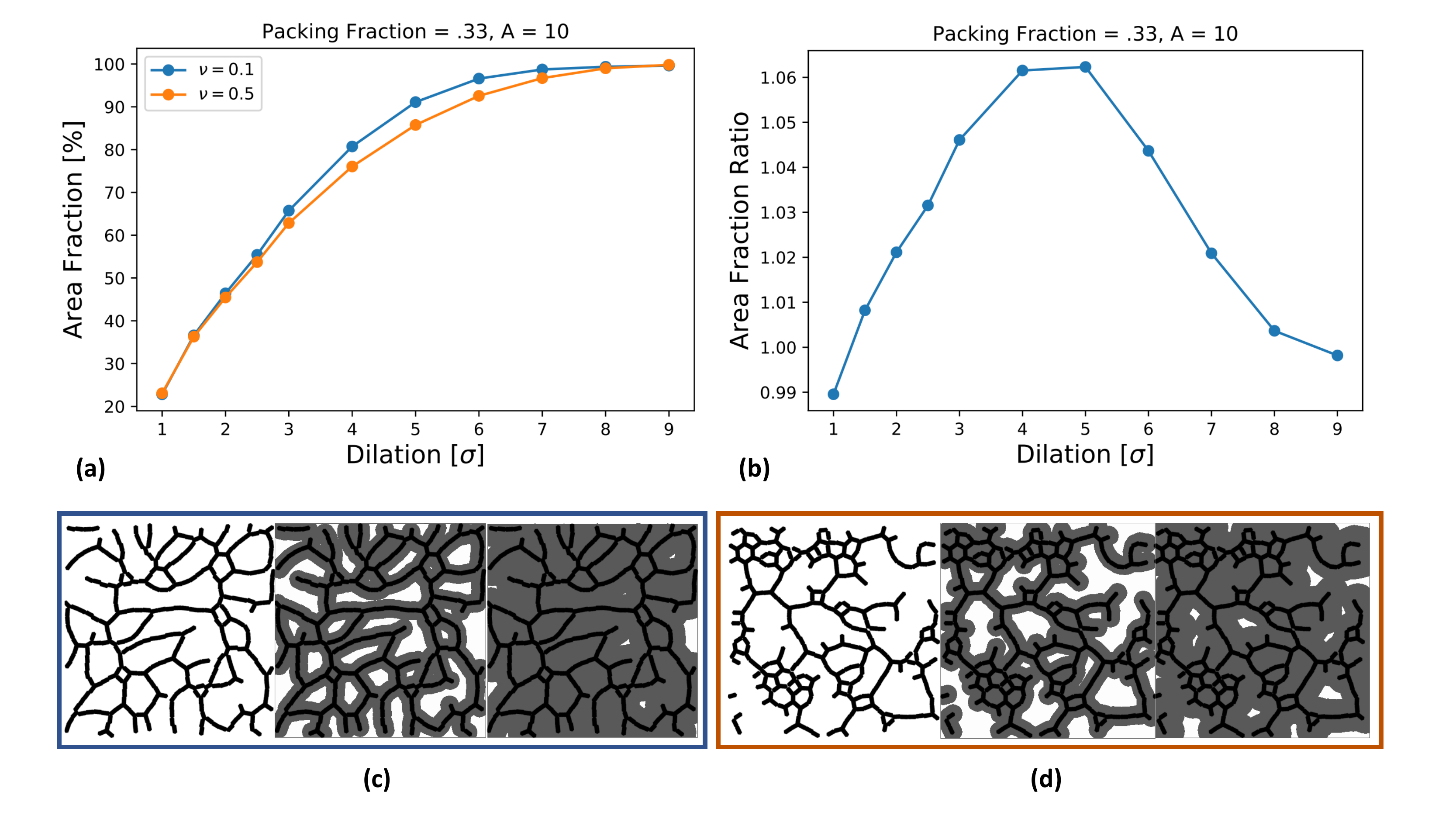}
		\caption{Substrate compressibility alters area coverage of  networks. (a)  Fraction of available area covered by simulated networks at $A = 10$ and $\phi = .33 (N=300)$, as the cell area is uniformly inflated by a dilation factor . $\nu = 0.1$ exhibits greater area for given dilation than the $\nu = 0.5$ case. This is due to the fact that higher values of $\nu$ produce  networks with more compact structures like junctions and 4-rings. These structures overlap one another when inflated unlike sparse networks with long branches. (b) Ratio of area coverage of $\nu = 0.1$ to $\nu = 0.5$ for $A = 10$ and $\phi = .33 (N=300)$. The plot increases sharply past unity then saturates to one at area limited dilation. (c) Visualizations of homogeneous dilation of a representative $\nu = 0.1$ network. (d) Visualizations of homogeneous dilation of a representative $\nu = 0.5$ network.}
		\label{fig:IrrigationArea} 
	\end{figure*}
	
The ability of a biological network to efficiently  cover space is crucial to deliver signals and materials.  Assuming the drainage area of each cell to be a dilation factor times the cell size, we analyze how the filling area of our networks scale with this cell dilation. We can then determine the density of interconnections within our networks. Supp. Fig.~\ref{fig:IrrigationArea} shows the filled area fraction of our networks as a function of homogeneous dilation factors. Realizations of these dilated networks at different dilation factors are shown as shaded regions in part c and d. This plot shows that the lower $\nu$ case increases area coverage as a function of dilation faster than the higher $\nu$ case. This reinforces the result in Fig. 6e as area fraction growth rate is maximized when cell overlap and proximity is minimized. Since higher values of Poisson's ratio produce networks with more compact structures of higher neighbor counts, its area coverage does not scale with dilation as strongly as the lower $\nu$ case. 

\section{Branch length analysis of simulation box and a larger composite box shows boundary errors are minimal}
\setcounter{equation}{0}
\begin{figure*}[!ht]  
		\centering
		\includegraphics[width=\textwidth]{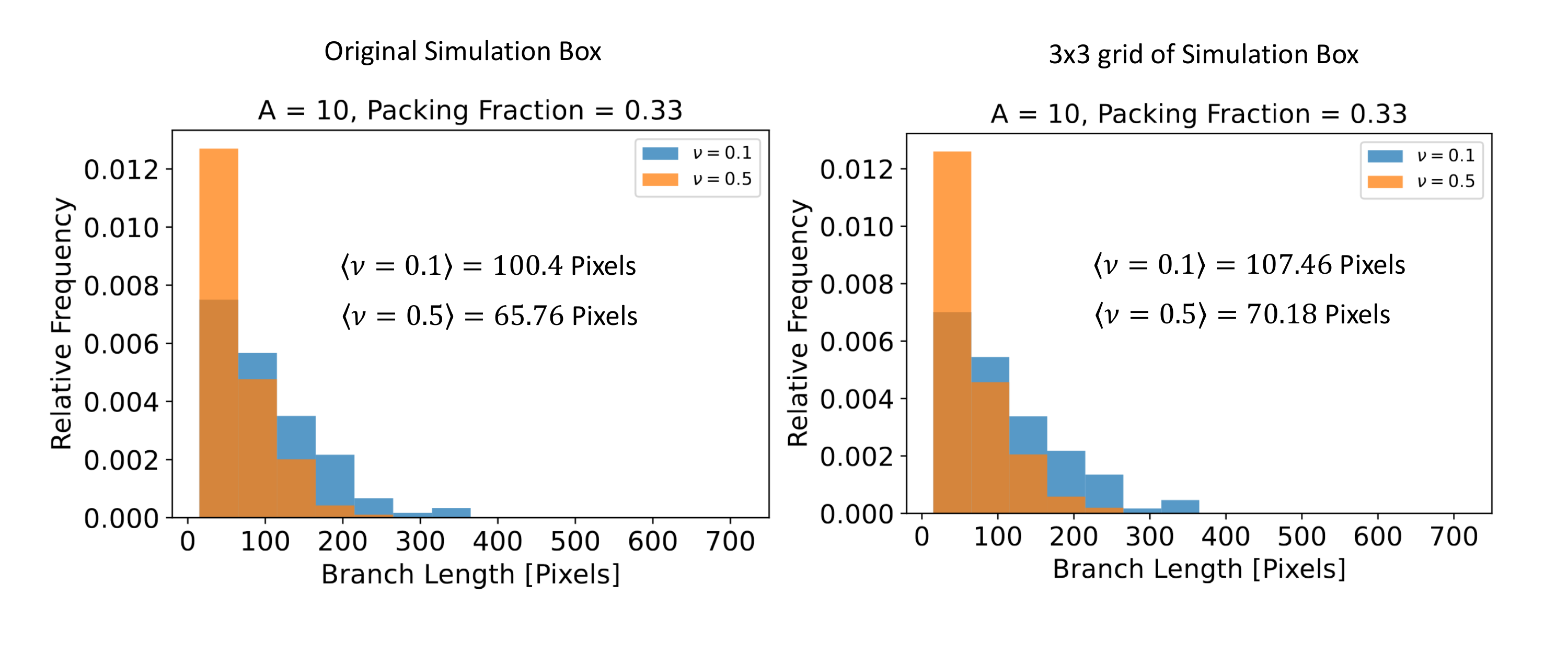}
		\caption{Branch Length histograms for both the original simulation box and a composite box reveal minimal boundary effects. (a) Branch length distribution for $A=10$ and $N=300$ with average branch length for our original simulation box. (b) Branch length distribution for $A=10$ and $N=300$ with average branch length when the original simulation box is made into an identical 3x3 grid of the simulation snapshots to reconstitute periodicity and study the effect of this boundary effect. Histograms are qualitatively similar and the error for the average branch lengths for both $\nu=0.1$ and $\nu=0.5$ is less than $10\%$. Thus, for computational feasibility of robustness studied in Fig.7d, we use the original simulation box snapshots.}
		\label{fig:BoxSize} 
	\end{figure*}
We utilize skeletonization in ImageJ for branch length, junction count, ring count, and robustness. ImageJ, however, does not account for the periodic boundary conditions by which cells interact. These boundary errors could lead to an incorrect analysis for the aforementioned metrics. To study the effect of these boundary errors, we report the branch length histograms and average branch lengths for both the original simulation box results and a box comprised of a 3x3 replication of the original simulation box in order to reconstitute periodicity and minimize boundary errors. Supp. Fig.~\ref{fig:BoxSize} shows that the error due to the boundary effects are minimal which lets us use the original box size for a computationally easier analysis.



\section{Simulation results for different choices of translational and rotational diffusivity}
\setcounter{equation}{0}
\label{Diffusion}
\begin{figure*}[!ht]  
		\centering
		\includegraphics[width=.9\textwidth]{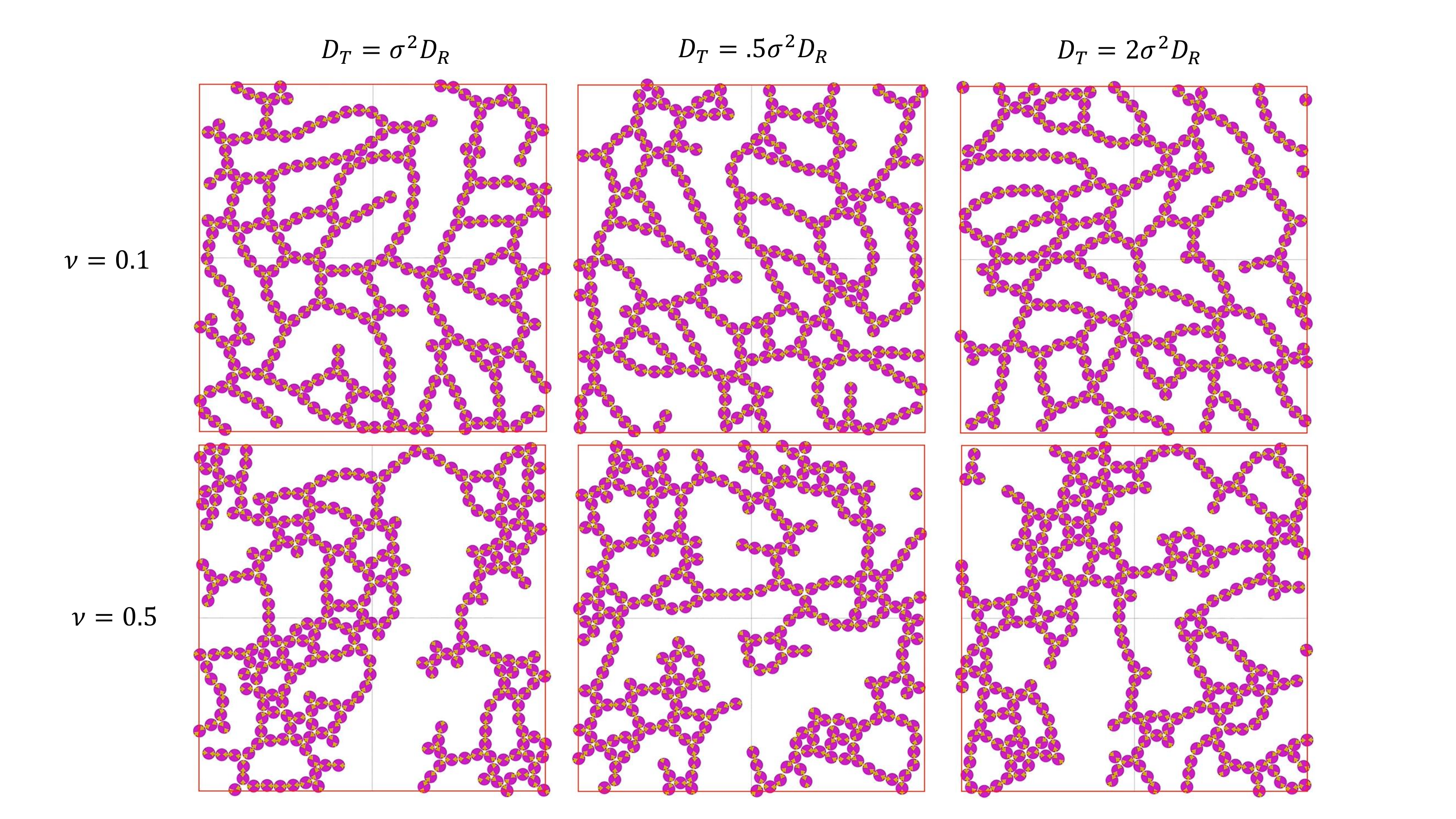}
		\caption{Network formation tendency robust to differing diffusion coefficients. (left) Networks formed from assumption stated in main text ($D_{T}=\sigma^{2} D_{R}$). (middle) Networks form when $D_{T}=.5\sigma^{2} D_{R}$. (right) Networks form when $D_{T}=2\sigma^{2} D_{R}$.}
		\label{fig:Diffusion} 
	\end{figure*}
While the results we present in the main text are for systems in which we choose the rotational and translational diffusivity to be proportional ($\sigma^{2} D_{R}=D_{T}$), this is not required to be the case for cells. The random movements of cells are caused by their internal physico-chemical activity, and the diffusivities are therefore not constrained by the fluctuation-dissipation theorem.  Supp. Fig. \ref{fig:Diffusion} shows that both $\nu=0.1$(top) and $\nu=0.5$(bottom) systems give rise to similar network configurations, whether the rotational diffusion is half or double its translational counterpart. This suggests that a different choice of rotational and translational diffusivity does not change the tendency of dipoles to form networks.  

\section{Simulation results for different choice of interaction cutoff range}
\setcounter{equation}{0}
\label{Range}
\begin{figure*}[!h]  
		\centering
		\includegraphics[width=.9\textwidth]{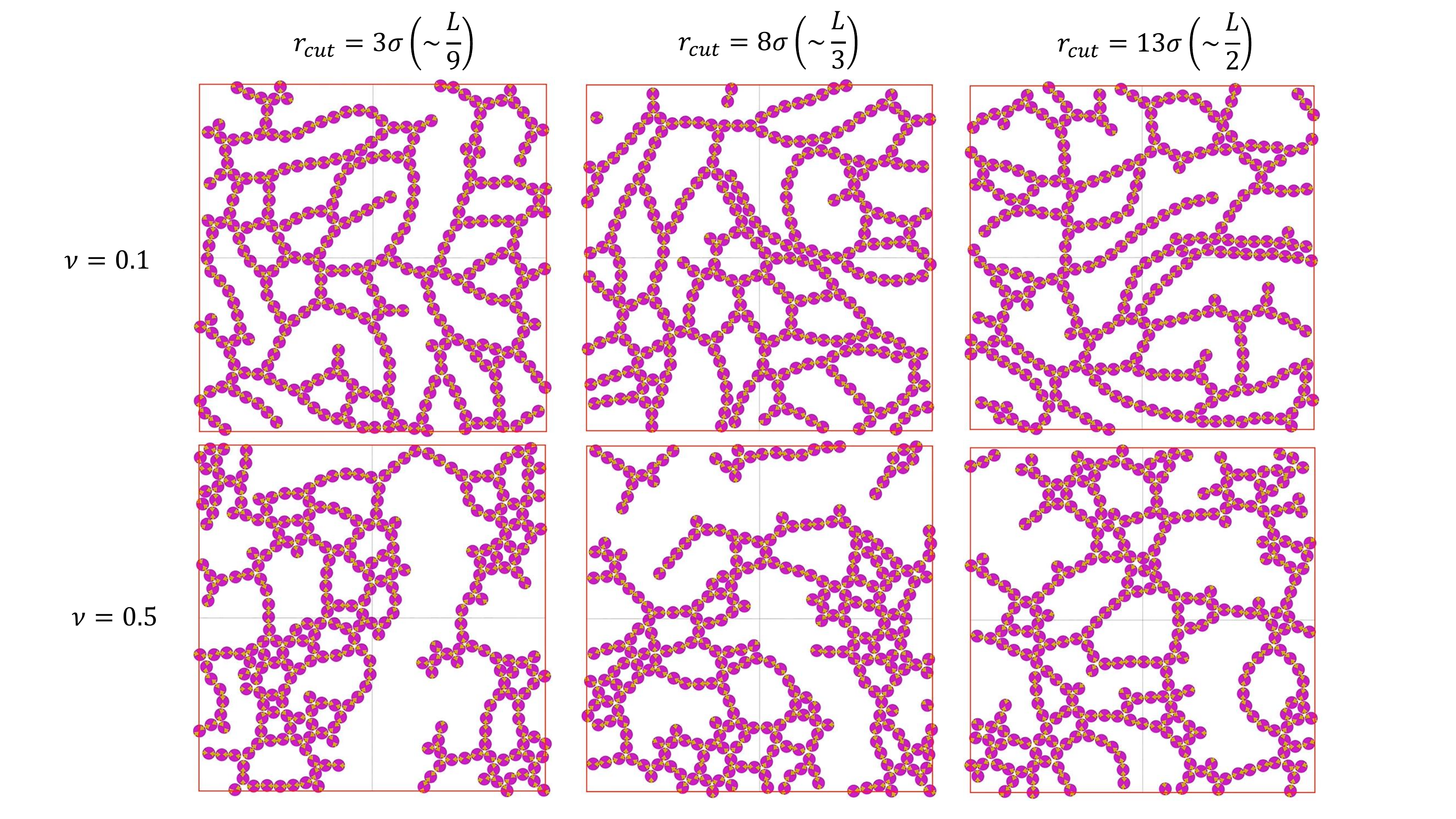}
		\caption{Network formation is qualitatively similar for longer cutoff ranges. (left) Networks formed from cutoff stated in main text ($r_{cut}=3\sigma$). (middle) Qualitatively similar network structures form when $r_{cut}=8\sigma$. (right) Qualitatively similar network structures form when $r_{cut}=13\sigma$.}
		\label{fig:Range} 
	\end{figure*}
In the main text, we cut off the long range elastic interactions at a value of $3 \sigma$, consistent with physiological limitations seen in cell culture experiments. Supp. Fig. \ref{fig:Range} shows utilizing much longer cutoff distances ($8 \sigma$ or $13 \sigma$) does not qualitatively change the assembly behavior. 

\section{Global ground states from on-lattice simulations}
\setcounter{equation}{0}
\begin{figure*}[!h]  
		\centering
		\includegraphics[width=\textwidth]{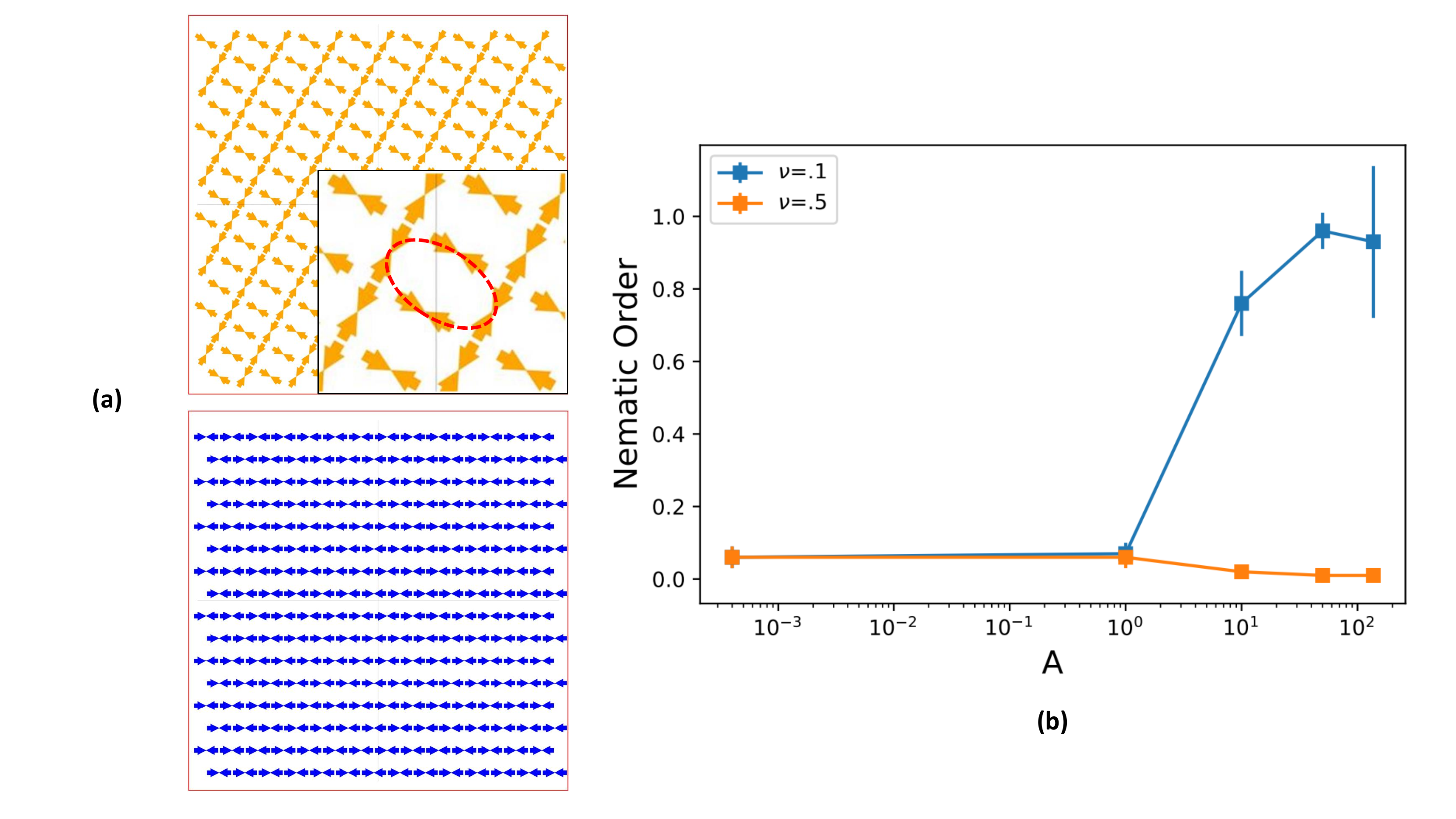}
		\caption{Final snapshots from on-lattice simulations of contractile for dipoles with only rotational freedom reveals ground state structures. Contractile force dipoles assume a global state of what previous literature has referred to as "4-rings" \cite{Bischofs2005} when $\nu=0.5$ (orange). Contractile force dipoles assume a global state of linear strings \cite{Bischofs2005} when $\nu=0.1$ (blue). (b) Nematic order parameter, $S \equiv 2\langle cos^{2}\theta \rangle - 1$ where $\theta$ is the difference between cell orientation and the average director, reveals as elastic interactions overcome noise, the $\nu=0.1$ system becomes entirely ordered (strings) while the $\nu=0.5$ system becomes entirely unordered or antiordered (4-rings).}
		\label{fig:OnLattice} 
	\end{figure*}
We can gain intuition on the common motifs in the network morphology, especially in the higher density situations, by  constraining the cell dipoles to a lattice. This corresponds to a situation where the dipoles have only rotational freedom, but the translation diffusion is very low leading to cells staying in their original positions.  Previous Monte-Carlo simulations of contractile force dipoles on a hexagonal lattice with rotational freedom showed the difference in ground state configurations between two different values of the Poisson's ratio of the elastic substrate - $\nu = 0.1$ and $\nu = 0.5$ \cite{Bischofs2006}. In Supp. Fig.~\ref{fig:OnLattice}, we reproduce these results using our Brownian dynamics simulations. Supp. Fig.~\ref{fig:OnLattice}a shows the ground state configurations of elastic dipoles where $\nu = 0.5$ in orange and $\nu = 0.1$ in blue. The dipoles tend to align globally at lower values of Poisson's ratio while at the higher Poisson's ratio,  the dipoles exhibit a mutual perpendicular alignment into ``4-rings''. The mutual alignment of the dipoles is measured by  a nematic order parameter magnitude, $ S = 2\langle\cos^{2} \theta^{\prime}\rangle - 1$. Supp. Fig.~\ref{fig:OnLattice}b shows this order parameter as a function of effective elastic interaction. At low values of $A$, noise destroys any order and coherent collective orientation. As $A$ increases, elastic interactions overcome stochastic effects and cells align with each other (become orthogonal to their neighbors) when $\nu = 0.1(0.5)$. Knowing the preferred configurations of these force dipoles in such a constrained case as a hexagonal lattice provides us the intuition with which we can understand the behavior of cells.









\end{document}